\documentclass[preprint,10pt,a4paper]{aastex}
\usepackage[colorlinks,urlcolor=blue,citecolor=blue,linkcolor=blue]{hyperref}
\usepackage{amsmath,amssymb}
\usepackage{graphicx}
\usepackage{multicol}
\usepackage[normalem]{ulem}

\usepackage[utf8]{inputenc}
\usepackage[T1]{fontenc}
\usepackage{textcomp}
\usepackage{amsmath}
\usepackage{amssymb,hyperref}
\usepackage{bbm}

 \usepackage{import}
 \usepackage{xifthen}
 \usepackage{pdfpages}
 \usepackage{transparent}

\newcommand{\B}{{\bf B}}

\newcommand\eg{\textit{e.g.,\ }}
\newcommand\cf{\textit{cf.\ }}

\begin{document}

\title{Magnetic topology  in  coupled  binaries, spin-orbital resonances,  and flares}
\author{Sergey A. Cherkis}
\affil{School of Mathematics, Institute for Advanced Study, Princeton NJ 08540, USA and\\
Department of Mathematics,  University of Arizona, 617 Santa Rita Ave., Tucson AZ 85721-0089, USA; cherkis@ias.edu}
\author{Maxim Lyutikov}
\affil{Department of Physics  and Astronomy, Purdue University,   525 Northwestern Avenue, West Lafayette, IN47907-2036, USA; lyutikov@purdue.edu}

\begin{abstract}
We consider topological configurations of the magnetically coupled spinning stellar binaries (e.g.,  merging neutron stars or interacting  star-planet  systems). 
We discuss conditions when the stellar spins and the orbital motion nearly `compensate' each other, leading to  very {\it slow} overall winding of the coupled magnetic fields; slowly winding configurations allow   gradual accumulation of magnetic energy,  that is eventually released in a flare when the instability threshold is reached.   We find that this  slow winding can be  global and/or local.    We describe  the topology of the relevant space  $\mathbb{F}=T^1S^2$ as the unit tangent bundle of the two-sphere   
and find conditions for  slowly winding configurations in terms of magnetic moments, spins and orbital momentum. These conditions become ambiguous near the 
topological  bifurcation points;  in certain cases they also depend on the relative  phases of the spin and orbital motions.       In the case of merging  magnetized neutron stars,      if one of the stars is a millisecond pulsar, spinning at  $\sim$ 10 msec, the global  resonance $\omega_1+\omega_2=  2 \Omega$ (spin-plus beat is two times  the orbital period)    occurs approximately a second   before the merger; the total energy of the flare can be as large as $10\%$ of the total magnetic energy,  producing bursts of luminosity $\sim    10^{44}$ erg s$^{-1}$.   Higher order local resonances may have similar powers, since the amount of involved magnetic flux tubes may be comparable to the total connected flux. 
           \end{abstract}

\maketitle

\section{Introduction}

Direct magnetospheric interaction occurs in binary Main Sequence stars \citep[\eg epsilon Lupi system,][]{2015MNRAS.454L...1S},  white dwarf binaries \citep{1983ASSL..101..155W,2017NatAs...1E..29B}, planetary systems \citep{2000ApJ...529.1031R,2021arXiv210405968A}, and has been suggested to occur between merging neutron stars \citep{2001MNRAS.322..695H,2011PhRvD..83l4035L,2013PhRvL.111f1105P,2018ApJ...869..130R,2019MNRAS.483.2766L,2020ApJ...893L...6M}. In the latter case it may lead to the precursor emission - production of an electromagnetic signal before the merger.

In the case of merging neutron stars it is expected  that the  {\it persistent} power of the EM precursor is not very high, and not likely to be detected by all-sky high energy monitors  \cite[][and \S \ref{merging}]{2001MNRAS.322..695H,2019MNRAS.483.2766L,2020ApJ...893L...6M}. Can the merging neutron stars produce flares that temporarily result in higher fluxes? 
 Stellar flares that release up to $10^7$ times more energy than the largest solar flare have  been detected from  main-sequence stars that host large planets  \citep[\eg][]{2000ApJ...529.1026S}.  \cite{2000ApJ...529.1031R} proposed that  super-flares are caused by magnetic reconnection between the primary star and  close-in Jovian planets.  Following this ideas, we will investigate possible appearance of flares in magnetically interacting stars, and neutron stars in particular.

\begin{figure}[!ht]
\centering
\includegraphics[width=.7\textwidth]{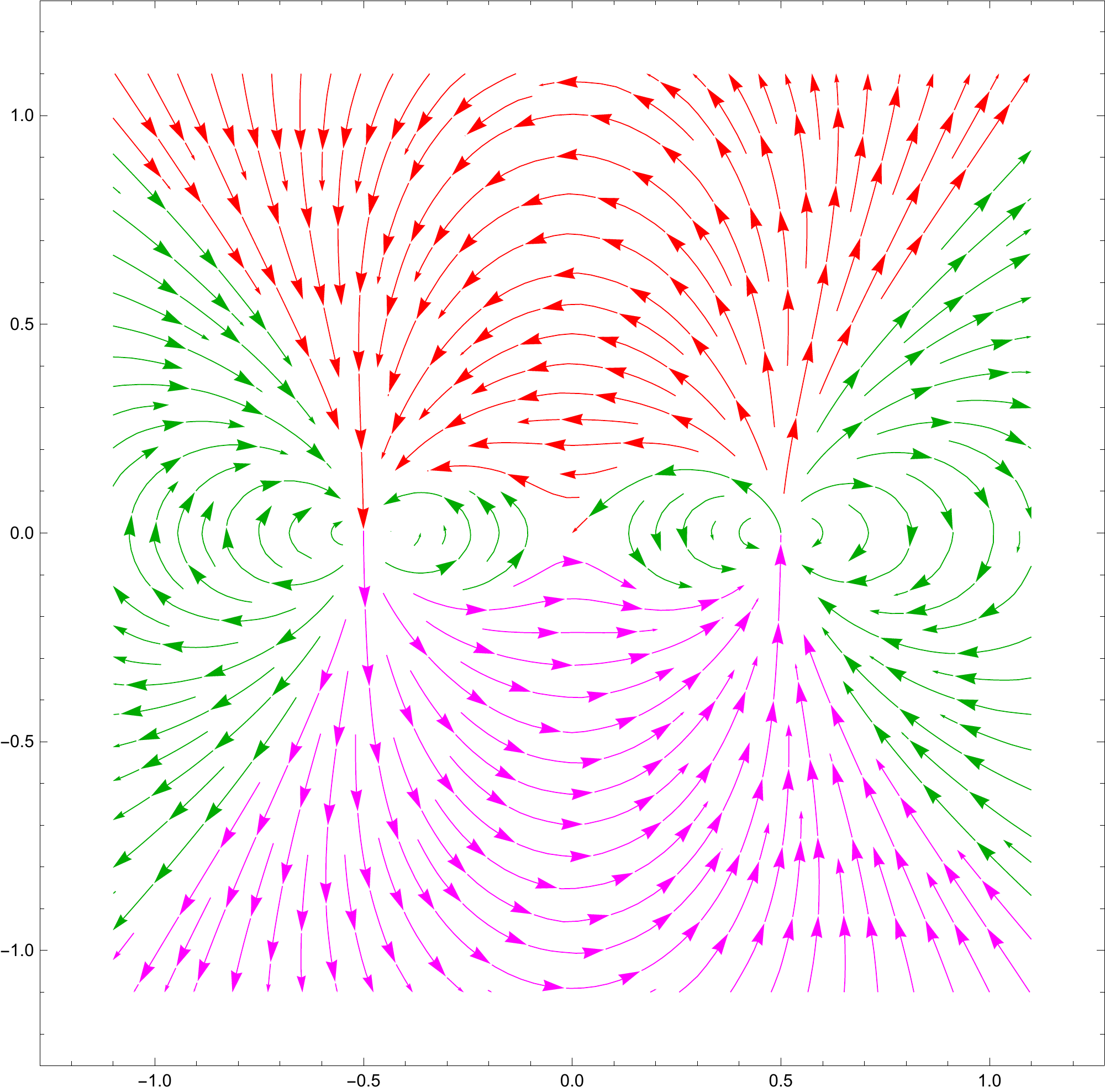}
\caption{Example of magnetically coupled stars. The stars are located at $x = \pm 0.5$ and have oppositely directed magnetic moments. The green  field lines end on the same star from which they originate and form region (i).  The red and magenta lines form region (ii), with red lines connecting the right star to the left one and the magenta lines connecting the left star to the right one.}
\label{open-close} 
\end{figure}

 A magnetosphere of magnetically interacting stars is expected to have three types of regions, Fig.~\ref{open-close}, formed by magnetic field lines that  (i) start and end on the same star; (ii)  provide magnetic coupling, and (iii) connect to infinity  \citep[such regions appear both due to the spin of each star and orbital motion][we ignore them here]{GJ}.
The common  part of the magnetosphere  (magnetically coupled region-ii) is twisted both by the relative spins of each companion, 
and  by the orbital motion of the companions. Of particular interest are quasi-stationary configurations of the interacting magnetospheres, when the twisting produced by the orbital rotation is (partially) compensated by the spin(s) of the binary.
In such cases we expect that the common magnetosphere is slowly wound/twisted by the combined effects of the components' spin and orbital motion -- as a result a fraction of the rotational energy is slowly stored in the magnetic field. After a system reaches some instability threshold the stored magnetic energy can be released in (possibly) observable flares. In contrast, in highly time-dependent configurations the energy release is expected to be nearly continuous - this results in smaller instantaneous luminosities and, in addition, quasi-steady sources are harder to detect observationally. 

In this paper we discuss the topology of magnetically interacting stars and identify (quasi)steady configurations, both global   (when the whole magnetosphere returns to an initial state),  and local (when only tubular neighborhoods of special magnetic field lines are untwisting). 

 The plan of the paper is as follows. In \S \ref{theconcept} we discuss qualitatively magnetic tube winding rate using the principle of braiding of present and future field lines. In \S \ref{sec:cases} we discuss
globally untwisting configurations. In \S \ref{Topology} we develop a mathematical description of field line windings and sue it to identify various possible resonances.  In \S \ref{merging} we discuss astrophysical applications of the model.

\section{The concept  of magnetic tube winding} 
\label{theconcept}

Consider two stars orbiting each other with orbital frequency $\Omega$. Consider distances much smaller than the (effective) light cylinder radius,  so that in the frame of the rotating stars the effects of line sweep-back are not important. It is expected that in all astrophysically important applications the surrounding can be described as plasma: even in the case of merging neutron stars, when little external plasma, the magnetospheres are filled with self-generated electron-positron plasma  \citep{GJ}.

 We are facing a complicated, time-dependent three-dimensional MHD problem (relativistic MHD in the case of merging neutron stars). 
 To get a physical insight, let us think in terms of magnetic flux tubes;  in nearly ideal plasma a magnetic flux tube has a clear physical interpretation due to the frozen-in condition. 
  Let us next construct a physical model of the magnetic fields with frozen in plasma, representing them as material objects: flux tubes. Each flux tube consists of nearby magnetic lines bundled together forming a tubular neighborhood of its central line.

  We are not interested in twisting of each field line by itself (which is ${\rm curl\, } \B$), but rather we are interested in the rate at which a magnetic tube's twist increases (its winding rate). That means comparing how  adjacent field lines and their images at a later time intertwine/braid around each other. Each field line can carry a current (can be twisted); this is not important. What we are after is the change in winding as measured by braiding of past and future magnetic lines. For example,  each hair in a braid can be twisted on its own, but the topology of a braid is determined by how adjacent hairs interweave around it relative to each other. 
Clearly, it takes {\it three} strands to define a braid (indeed, the hair braid needs at least three strands of hair). In our case one of the strands will be the central line, another -- the nearby line from the past, and the third -- that same nearby line from the future. 
    In more detail: since we are interested in the change of twisting of the flux tubes, it suffices to track its central line   and one of the nearby lines (as a reference).  This pair of lines can be thought of as a ribbon, with differently colored sides. 
    Say, the central line is blue and its near by line at the initial moment is red.  
    Whenever at some later moment the central line happens to be sufficiently close to its original shape, we have the nearby line at that moment provide the third strand we need. Color the nearby line at that later moment is green. Thus the braid consists of the blue central line strand (initial same as final), the red initial  nearby line strand and the green final nearby line strand.  Then, at that moment we can measure the winding of the final (green) nearby line around the (blue) central line relative to the (red) initial nearby line. The winding rate is that winding angle divided by the time increment. 

    Of course, whenever the central line has no inflection points and, therefore, has a Frenet triad (consisting of a unit tangent, a principal normal, and their vector product), then there is a good notion of twisting: in this case the twisting of the magnetic tube is the angle by which a nearby line winds relative to the principal normal as we traverse the line.  This is usually called twisting and denoted by $Tw$ \cite{moffatt_dormy_2019}.  What we study is the rate of change in twisting $\frac{d }{dt}Tw$, which we call {\em winding rate}. This rate is well defined even when $Tw$ is not (when the curve develops inflection points), as described above. 

    Intuitively, one might expect a relation of the above notion of magnetic tube twisting and helicity, since helicity measures average magnetic field self-linking \cite[]{Arnold73,Arnold74}. Indeed, there is some relation, though, not as direct as one would like; namely, helicity is a sum of writhe and twist \cite[Sec.~2.10]{moffatt_dormy_2019}.   Our focus is on twist of the magnetic line, or, rather, on its rate of change. Thus our approach cannot be reduced to the  commonly discussed  magnetic helicity.

\section{Globally non-winding magnetic configurations of orbiting stars}
\label{sec:cases}

\subsection{The basic 2:1 resonance}

As discussed above, the key point to production of observable  flares is the establishment of quasi-steady magnetic configuration, whereby the magnetic energy is  slowly stored in the magnetospheres  and later released in a sudden flare. 

There are  several basic cases where we expect that the  whole interacting magnetospheres periodically return to their initial state. Consider first the case of magnetic moments parallel to the $z$-axis (one aligned, one counter-aligned, so that the two stars are magnetically connected) and spins also (anti)/parallel to $z$. Case-I  is a fully locked case: $\omega_1=\omega_2=\Omega$. 
In the rotating frame this corresponds to $\omega_1'=\omega_2' =0 $ (dashed quantities are measured in the corotating frame). Neutron stars are not expected to be tidally locked \citep{1992ApJ...400..175B}, so this case is unlikely to be realized and, consequently, of no interest for us.

\begin{figure}[!ht]
\centering
\includegraphics[width=.99\textwidth]{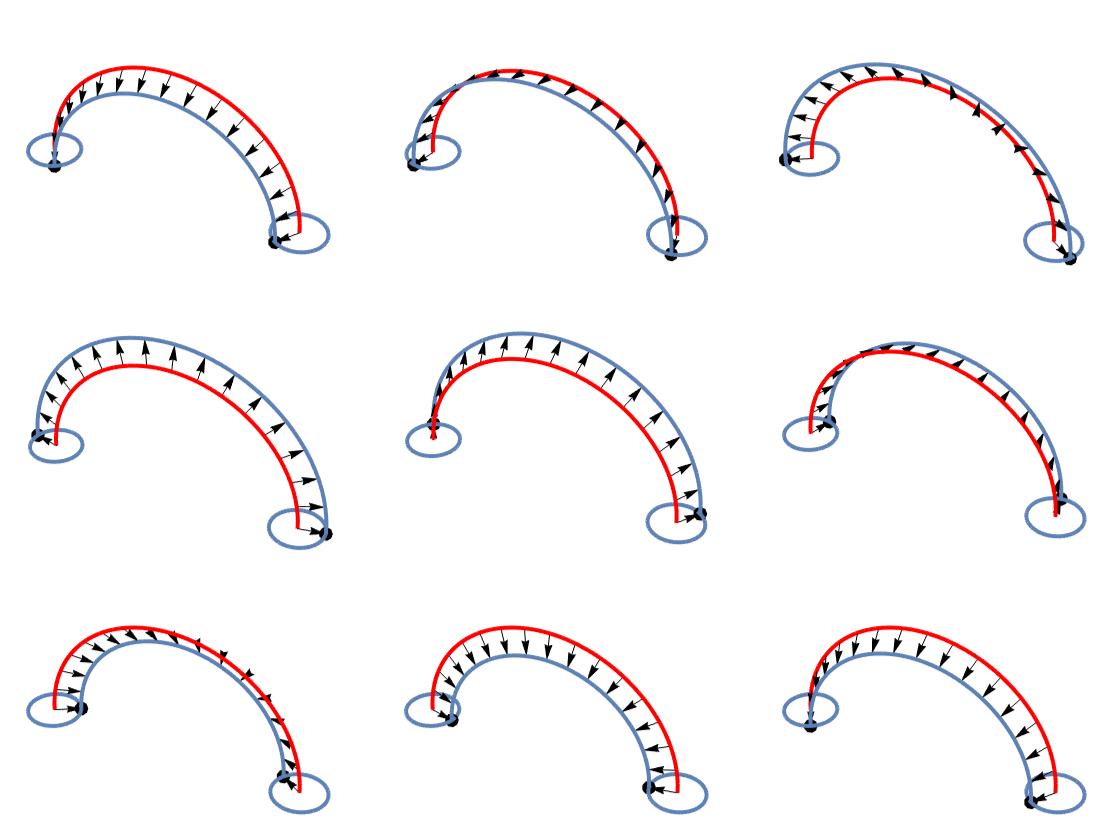}
\caption{ Basic configuration for Case-II:  magnetically couples neutron stars in the corotating frame, where $\omega_1'= - \omega_2'$. The two neutron star are represented by two disks connected by magnetic fields. Since we are not interested in the detailed dynamics here, but only in topological properties, we can approximate magnetic field lines are semi-circles attached to particular points on the disks. Arrows indicate the normal field. After a full rotation the system comes to the original state.
Dots on the edges of the disks and differently colored flux lines help visually trace the evolution. 
 }
\label{twisted} 
\end{figure}

\begin{figure}[!ht]
\vskip -.15truein
\includegraphics[width=.99\linewidth]{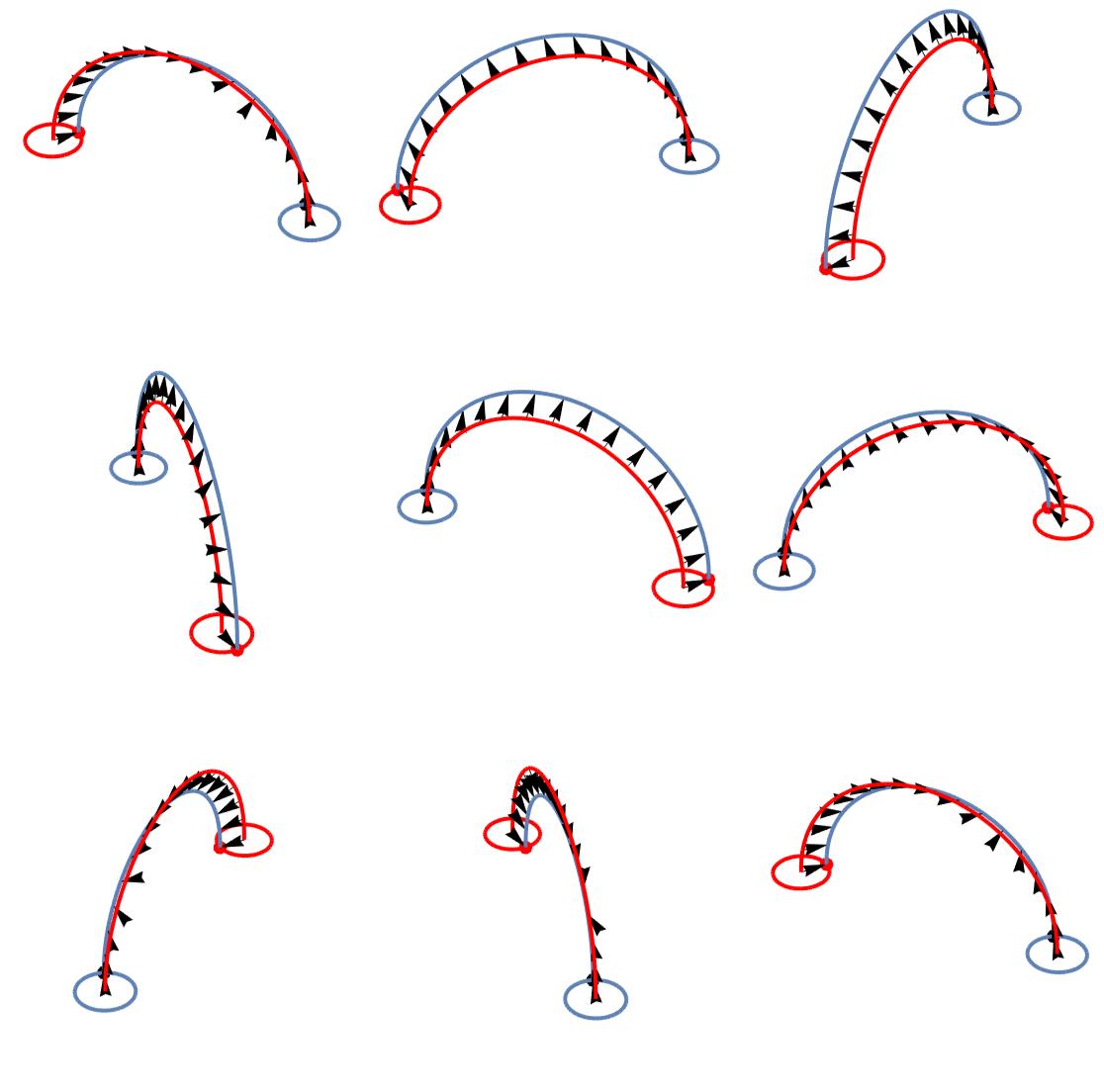}
\caption{ A basic example of the  2:1 spin-orbital resonance. In this example the blue disk is non-rotating.}
\label{New-Final}
\end{figure}

\begin{figure}[!ht]
\vskip -.15truein
\includegraphics[width=.99\linewidth]{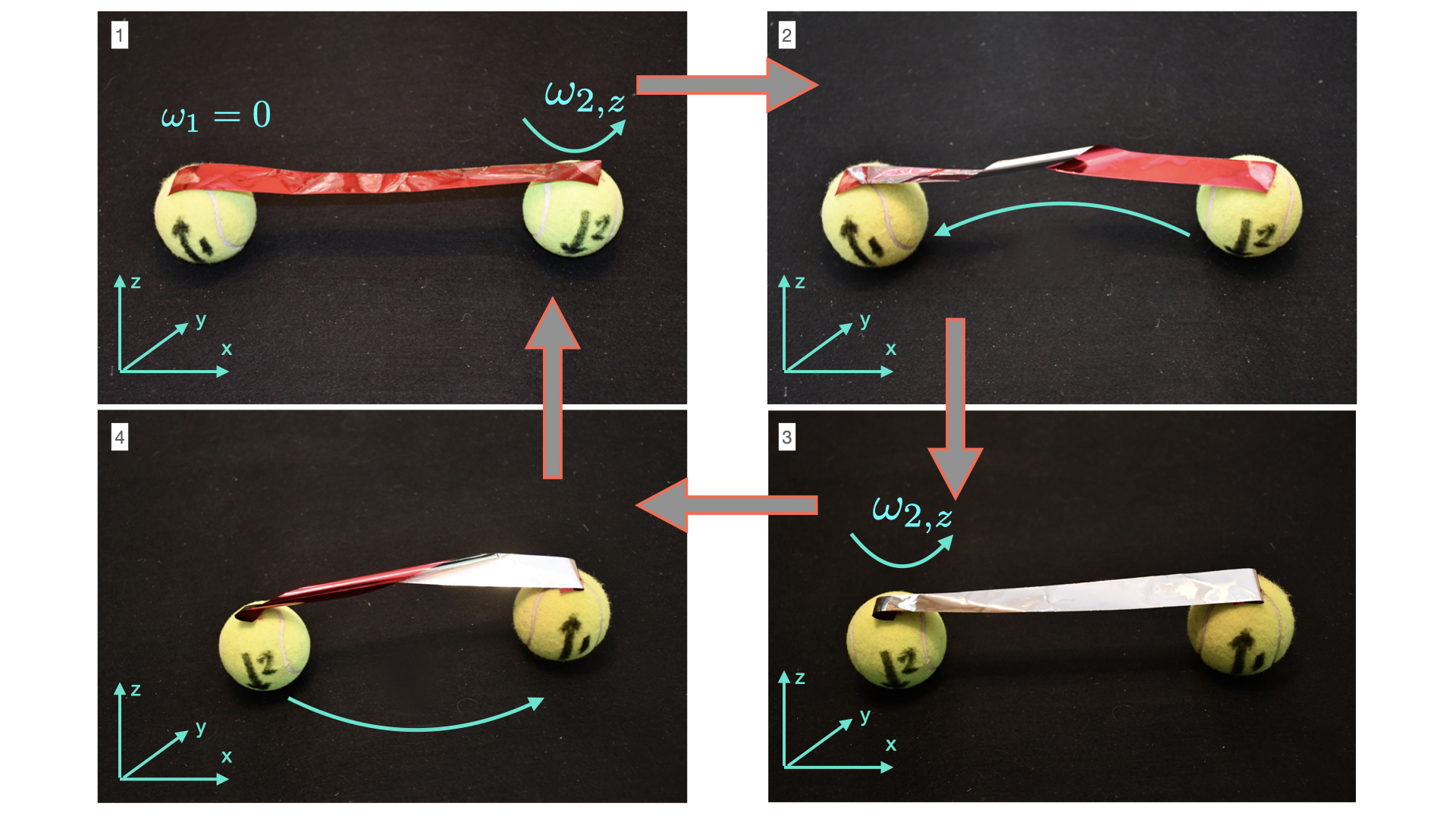}
\caption{ An example of the  2:1 spin-orbital resonance (Case-II). The tennis balls represent  two neutron stars. Black arrow indicates the direction of the magnetic moment.  Two hemispheres (East and West) are clearly identified:  this helps keep track of the spin phase. The ribbon is the magnetic field connecting the stars; magnetic moments are counter-aligned, so that there is a strong magnetic coupling between the stars. The star $1$ is non-spinning. The star $2$ first makes a full spin around $z$ axis, which is orthogonal to the orbital plane, (panel 2), then half a rotation around star $1$ (panel 3), then another  full spin    (panel 4), and other half orbital rotation brings it back to the initial configuration  (panel 1).  Such orbital untwisting of the common magnetosphere twisted by the spins occurs generically for many spins, orbital angular momentum and magnetic moments directions.}
\label{New-Final2}
\end{figure}

Case-II is counter-aligned spins in the corotating frame  $\omega_1' = - \omega_2'$ (and arbitrary $\Omega$), Fig. \ref{twisted}. In this case in the rotating frame the configuration returns to the initial state. In the observer frame this corresponds to the case when  the frequency of  beat-plus of spins equals two times the orbital frequency, Figs. \ref{New-Final} and  \ref{New-Final2},
\begin{equation}
\omega_1+\omega_2 = 2 \Omega.
\label{2to1}
\end{equation}
Importantly, the  condition (\ref{2to1}) applies to the beat-plus frequency $\omega_1+\omega_2 $, not each individual frequency separately. 
 We call this 2:1 resonance: the sum of spins is two times the orbital frequency, Fig. \ref{New-Final}.
This is the Dirac belt configuration. In the case of merging neutron stars the changing orbital frequency may, at some point, become two times the sum of spins.

To the best of our knowledge the 2:1 spin-orbital resonance  has not been applied to magnetic fields of interacting binaries.  
In general physics such twisting arrangement is known under the names of 
 ``Dirac's belt''  or  ``Feynman arrow'' (also related to a ``Plate trick''). 



\subsection{Variants of the 2:1 resonance, phases and bifurcations}

\begin{figure}[!ht]
	\centering
	\begin{tabular}[c]{l}
\includegraphics[width=\textwidth]{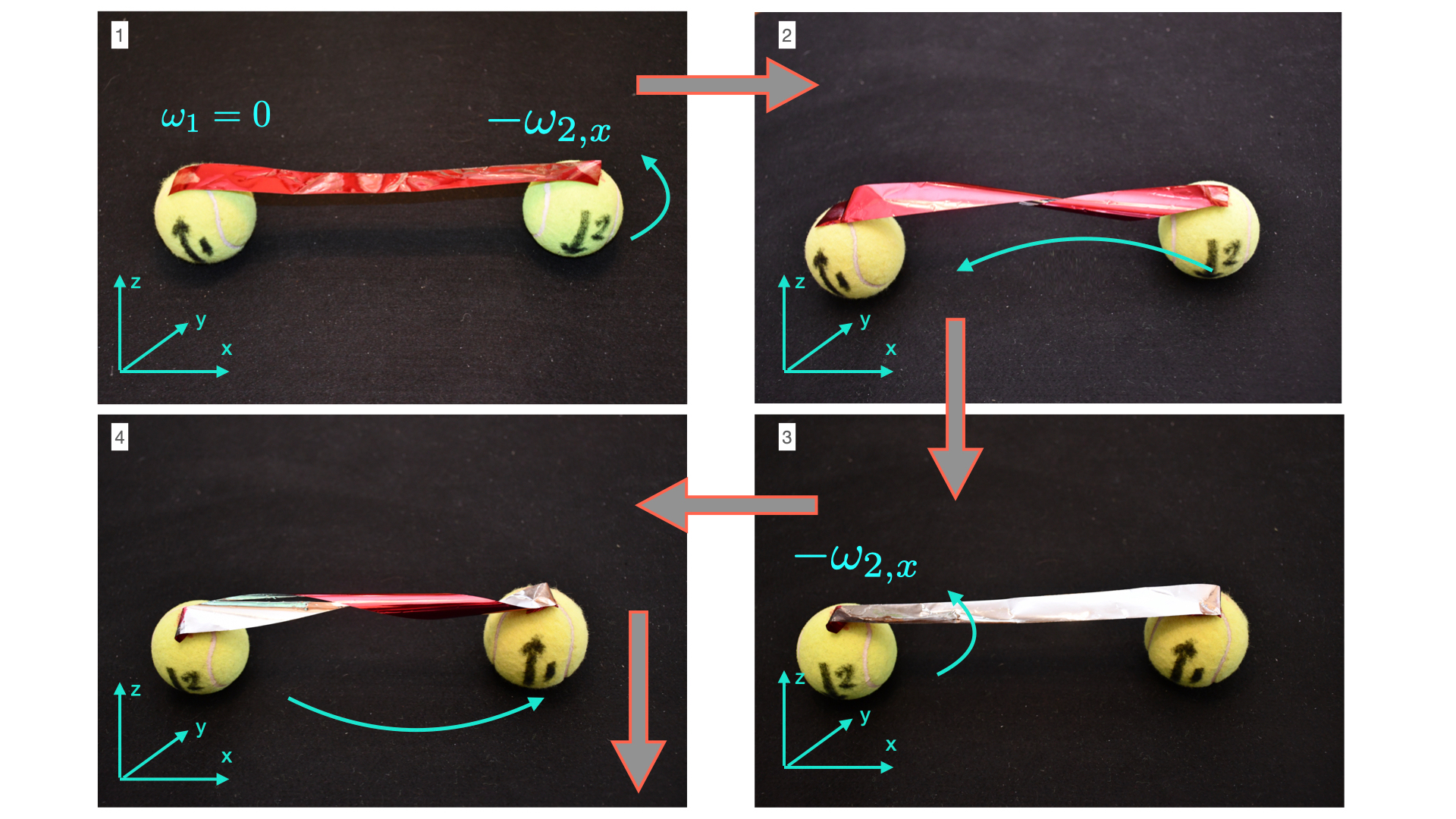}
\\
		\hspace{0.06\textwidth}\includegraphics[width=.42\textwidth]{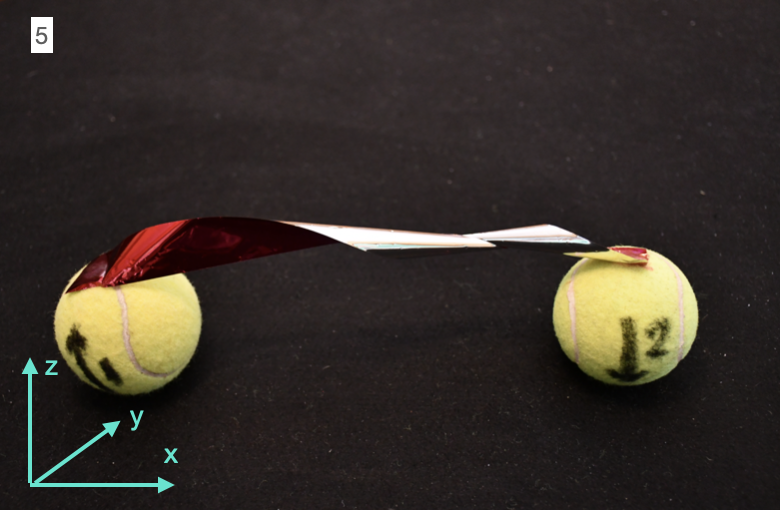}
	\end{tabular}
\caption{Second spin along the $x$-axis, with the second star undergoing two rotations during one orbital period. In this case the second star make one spin, then half an orbit, then another full spin, and then another half an orbit, the configuration {\it does not unwind},  \cf, Fig. \protect\ref{BasicX-unwind}.}
\label{Bifurcation-X-notunwind} 
\end{figure}

\begin{figure}[!ht]
\includegraphics[width=.99\textwidth]{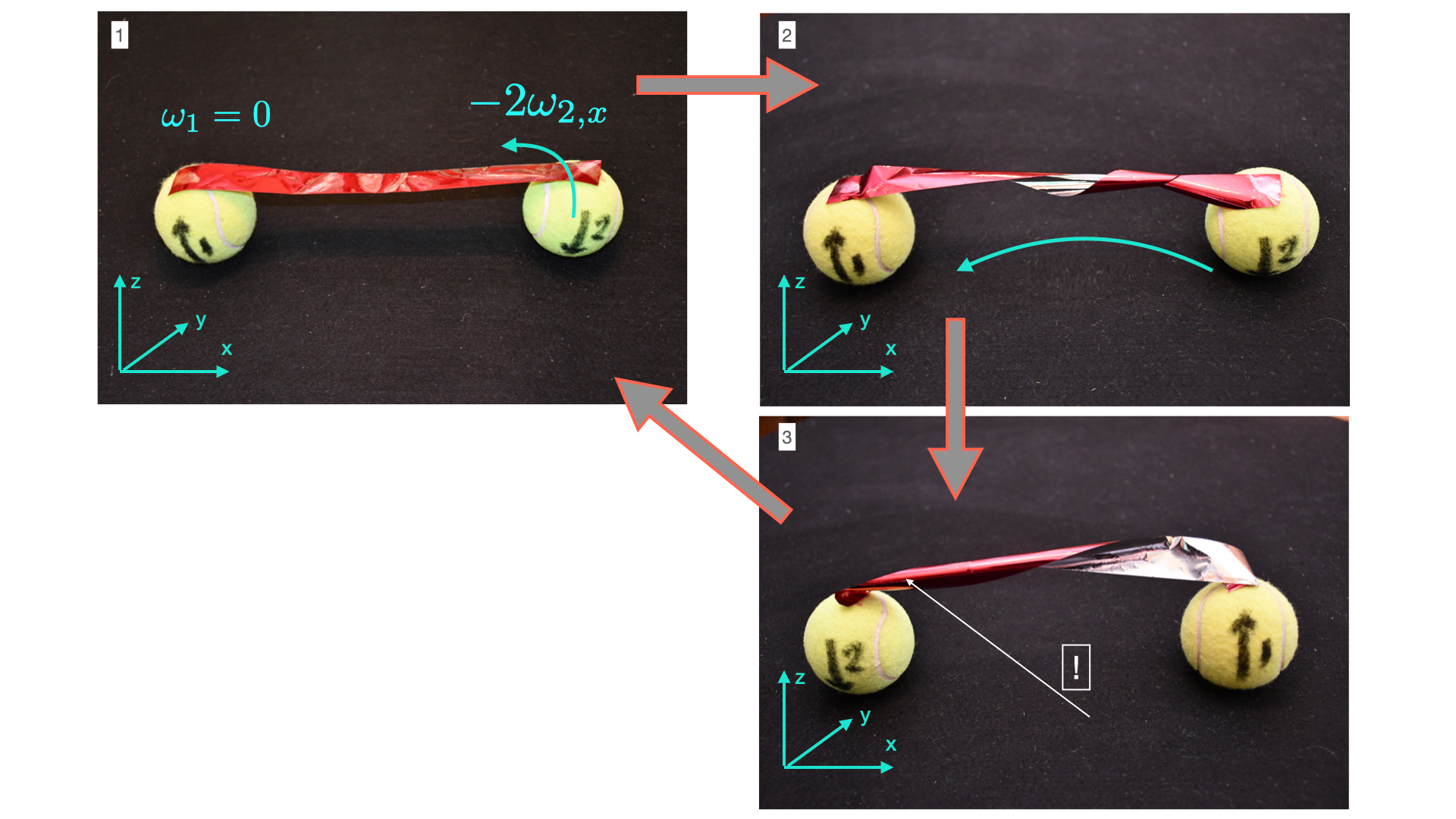}
\caption{Second spin along the $x$-axis, with the second star undergoing two rotations during one orbital period. In this case the second star make two spin, then  a full  orbit, the configuration {\it does  unwind},  \cf, Fig. \protect\ref{Bifurcation-Y-notunwind}}
\label{BasicX-unwind} 
\end{figure}

\begin{figure}[!ht]
\includegraphics[width=.99\textwidth]{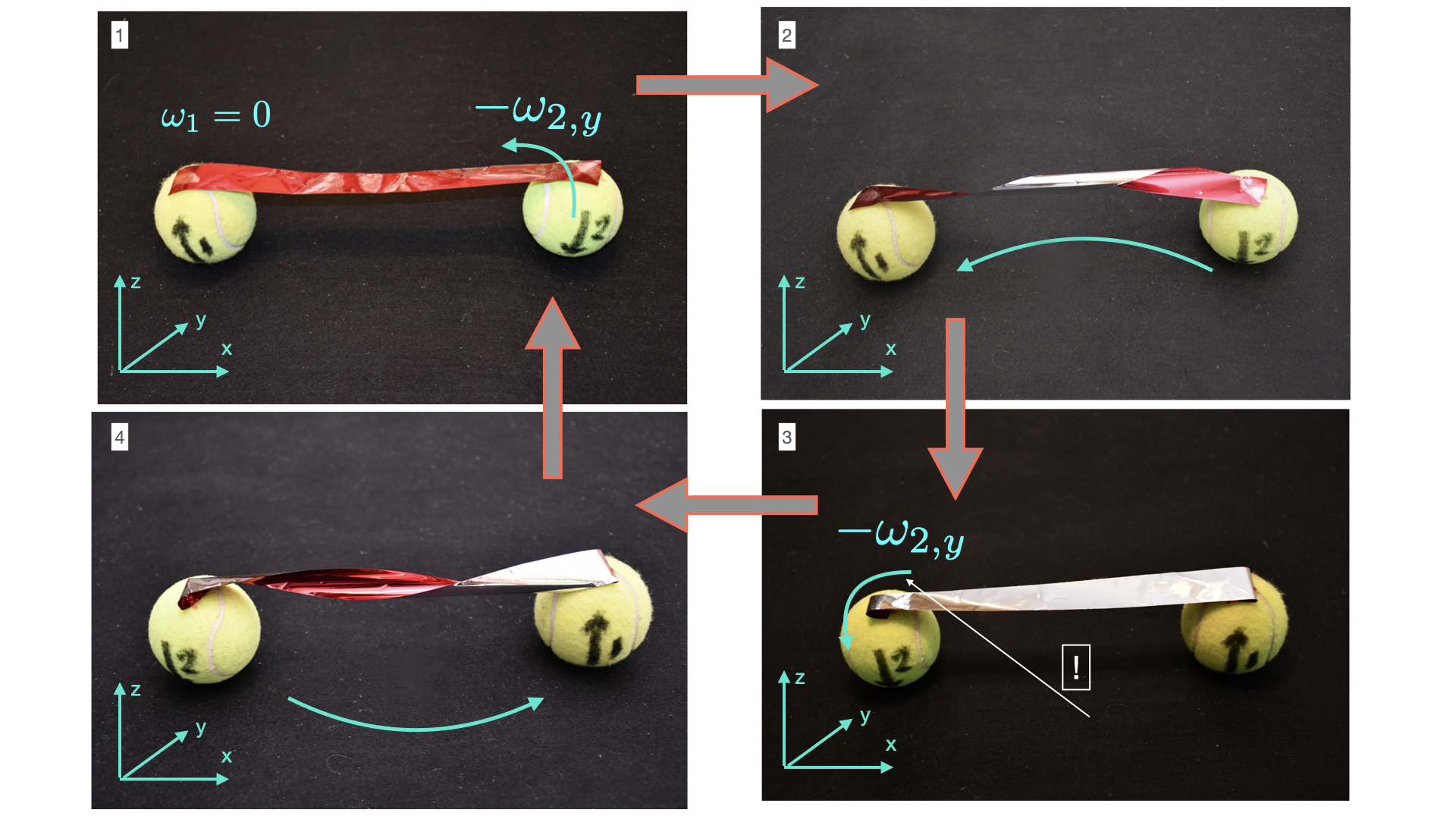}
\caption{Second spin along the $y$-axis. First the second star makes a full spin, with the  spin direction mostly along $y$ axis, but slightly misaligned in {\it positive } $x$  direction. After half an orbital turn the second star  makes a full spin, with the  spin direction mostly along $y$ axis, but slightly misaligned in {\it negative } $x$ direction. Configuration unwinds.    }
\label{Bifurcation-Y-unwind} 
\end{figure}

\begin{figure}[!ht]
	\centering
	\begin{tabular}[l]{l}
\includegraphics[width=\textwidth]{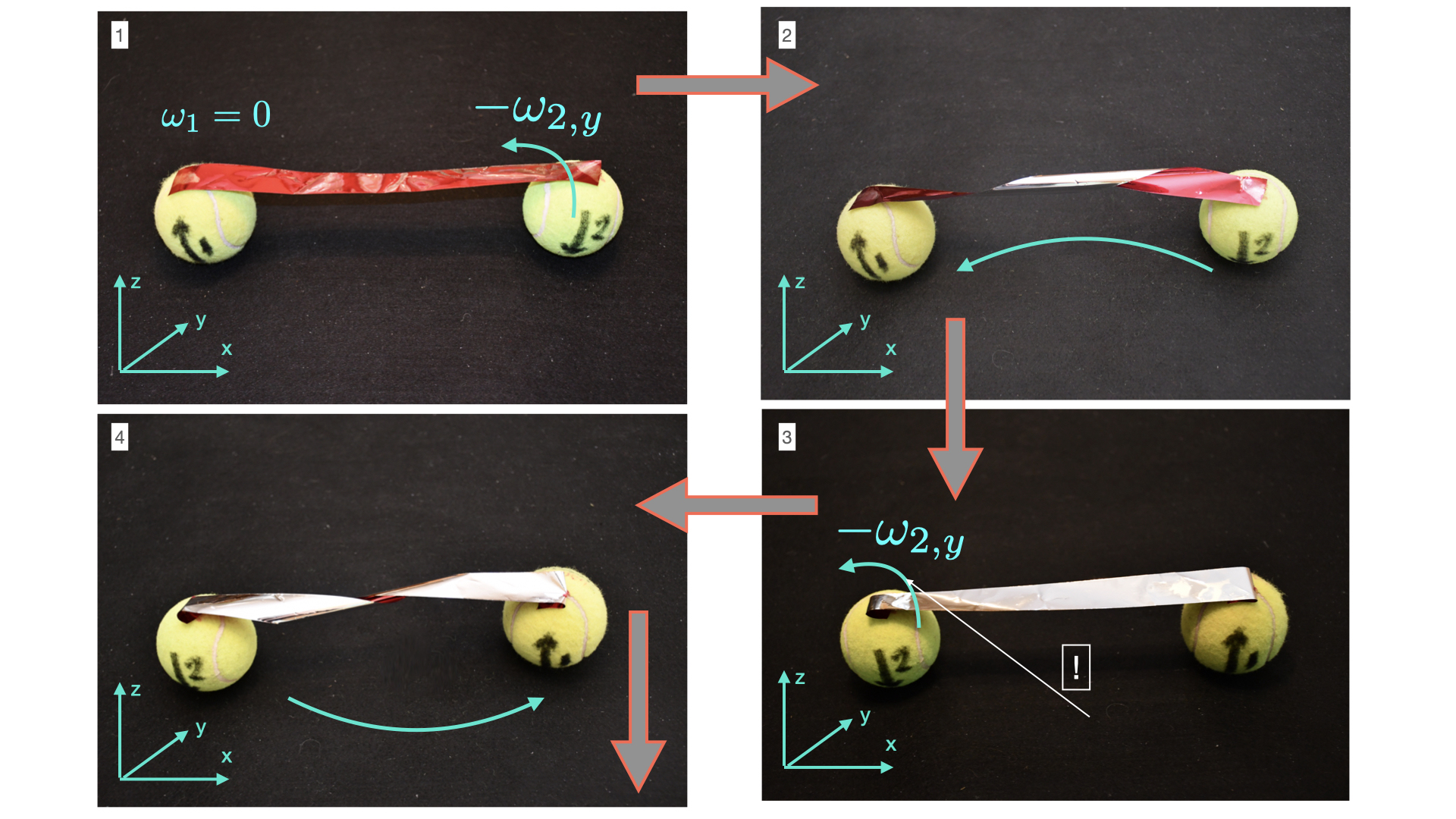}
	\\
	\hspace{0.07\textwidth}
\includegraphics[width=0.41\textwidth]{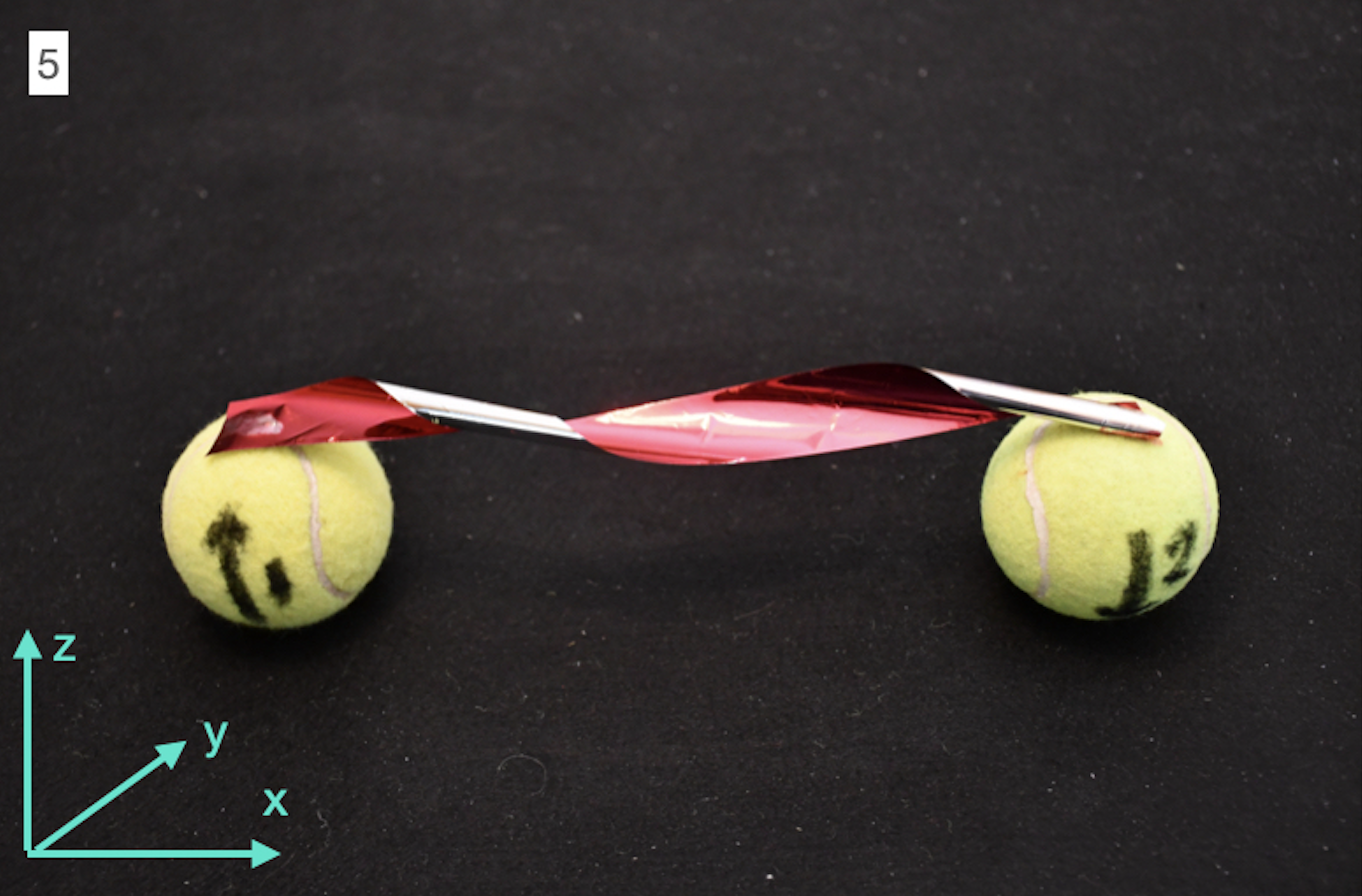}
        \end{tabular}
\caption{Second spin along the $y$-axis. First the second star makes a full spin, with the  spin direction mostly along $y$ axis, but slightly misaligned in {\it positive } $x$  direction. After half an orbital turn the second star  makes a full spin, with the  spin direction mostly along $y$ axis, but slightly misaligned also  in {\it positive } $x$ direction. Configuration  does not unwind.    }
\label{Bifurcation-Y-notunwind} 
\end{figure}

In addition to the basic  scenarios considered  above, there is a number of more complicated cases.  As we are about to demonstrate, the 2:1 
resonance  is more generic than the aligned case - it occurs for a wide variety of directions of  spins and orbital axes and orientations of the intrinsic magnetic  dipole moments. 
We shall now describe three indicative cases, illustrating them in Figs.~\ref{New-Final2}-\ref{Bifurcation-Y-notunwind} using table-top demonstrations.  In the next section, we give a more economical description of the relevant geometry, and use it to revisit these illustrative cases again with better insight. 

We number the two stars, let the first star not rotate at all, and, for definiteness, in all cases let the magnetic moments be counter-aligned along the $z$-axis at the initial moment (so that there is a strong magnetic coupling).   
Consider first when the spin of the second star is along the $x$ axis. Curiously, if the second star make one spin rotation first, then half an orbit, then another full spin rotation, and then another half an orbit, the configuration {\it does not unwind},  Fig. \ref{Bifurcation-X-notunwind}.
If, instead, the second star first makes two spin rotations, and then a full orbital rotation,  the configuration {\it doest unwind},  Fig. \ref{BasicX-unwind}. This example illustrates that the significance of the relative phase and angles. 

\begin{figure}[!ht]
\centering
\includegraphics[width=.99\textwidth]{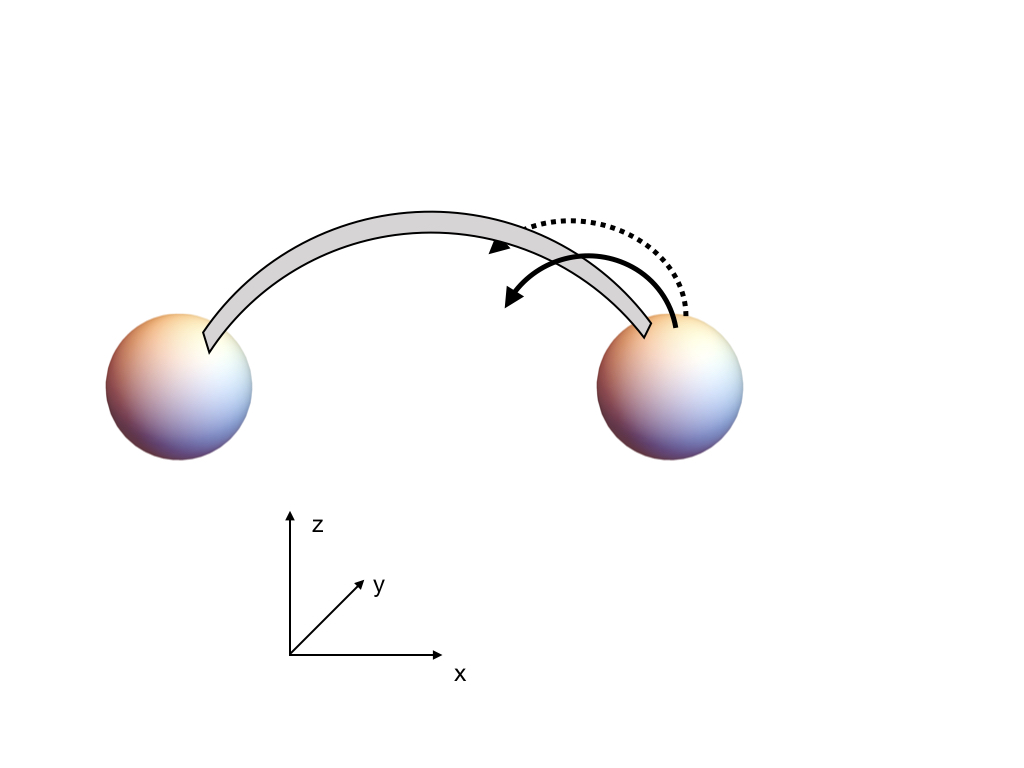}
\caption{Topological bifurcation. The star on the right spins approximately along $-y$ axis (which is into the board). Small variations of the spin direction would result in different topological behaviors:  one  not resulting in winding,  another resulting in winding.
 }
\label{bifucation} 
\end{figure}

Next, consider the case of spin of the second star along the $y$ axis. This case displays another property: topological bifurcation. In example in Fig.  \ref{Bifurcation-Y-unwind} the configuration unwinds, while in nearly equivalent case Fig. \ref{Bifurcation-Y-notunwind} it does not. Fig. \ref{bifucation} illustrates this behavior. 

\section{Topology of field winding}
\label{Topology}

In addition to three globally untwisting structures (fully locked, same counter-aligned spins, and 2:1 beat-plus resonance) there are other configurations that untwist locally. We study conditions for such untwisting next. To this end we first set up the geometric framework facilitating our search for non-winding configurations.

Our goal of studying the rate at which magnetic tubes are twisting is intimately related to topology.  With this in mind we discuss two convenient ways of thinking about magnetic tubes and then describe the topology of the relevant space, that can be viewed either as the rotation group $\mathbb{F}=SO(3)=\mathbb{R}P^3$ (Sec.~\ref{sec:SO3}) or as the unit tangent bundle of a two-sphere $\mathbb{F}=T^1S^2$ (Sec.~\ref{sec:F}).

\subsection{A magnetic tube}
Let us start with a convenient mathematical description of a magnetic tube, it is a sufficiently small tubular neighborhood of some central magnetic line.  It is formed by a smooth family of nearby magnetic lines parameterized by any transverse disk.  
(Mathematically, following the lines gives a diffeomorphism between any two transverse disks.)

\begin{figure}[!ht]
    \centering
	\def\svgwidth{1\columnwidth}
\begingroup%
  \makeatletter%
  \providecommand\color[2][]{%
    \errmessage{(Inkscape) Color is used for the text in Inkscape, but the package 'color.sty' is not loaded}%
    \renewcommand\color[2][]{}%
  }%
  \providecommand\transparent[1]{%
    \errmessage{(Inkscape) Transparency is used (non-zero) for the text in Inkscape, but the package 'transparent.sty' is not loaded}%
    \renewcommand\transparent[1]{}%
  }%
  \providecommand\rotatebox[2]{#2}%
  \newcommand*\fsize{\dimexpr\f@size pt\relax}%
  \newcommand*\lineheight[1]{\fontsize{\fsize}{#1\fsize}\selectfont}%
  \ifx\svgwidth\undefined%
    \setlength{\unitlength}{680.31496063bp}%
    \ifx\svgscale\undefined%
      \relax%
    \else%
      \setlength{\unitlength}{\unitlength * \real{\svgscale}}%
    \fi%
  \else%
    \setlength{\unitlength}{\svgwidth}%
  \fi%
  \global\let\svgwidth\undefined%
  \global\let\svgscale\undefined%
  \makeatother%
  \begin{picture}(1,0.5)%
    \lineheight{1}%
    \setlength\tabcolsep{0pt}%
    \put(0,0){\includegraphics[width=\unitlength,page=1]{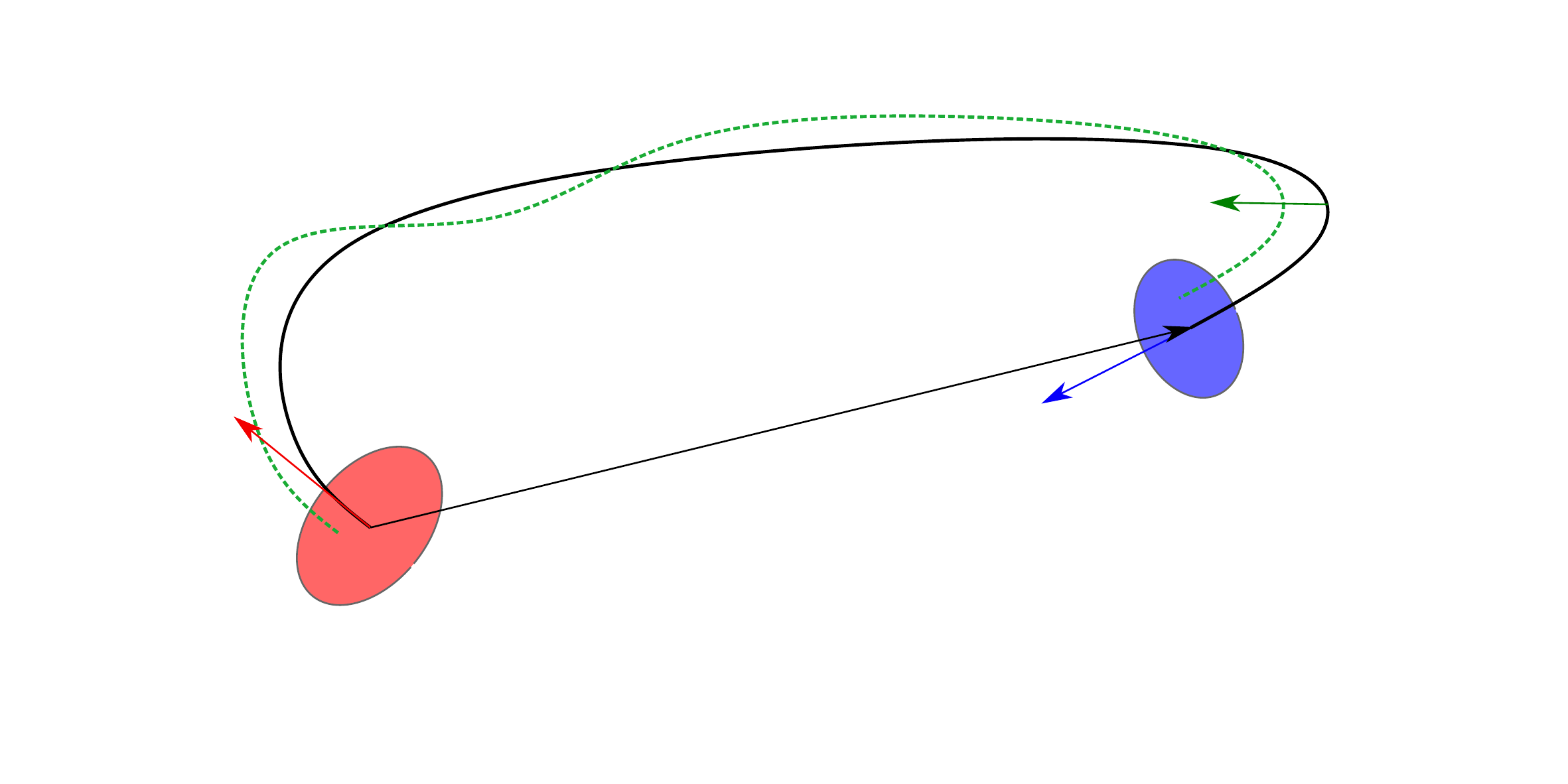}}%
    \put(0.47553129,0.24833234){\color[rgb]{0,0,0}\makebox(0,0)[lt]{\lineheight{1.25}\smash{\begin{tabular}[t]{l}$\vec{R}$\end{tabular}}}}%
    \put(0.67961349,0.22394168){\color[rgb]{0,0,0}\makebox(0,0)[lt]{\lineheight{1.25}\smash{\begin{tabular}[t]{l}$\vec{\mu}_2$\end{tabular}}}}%
    \put(0.12417082,0.22505035){\color[rgb]{0,0,0}\makebox(0,0)[lt]{\lineheight{1.25}\smash{\begin{tabular}[t]{l}$\vec{\mu}_1$\end{tabular}}}}%
    \put(0.74704672,0.36773586){\color[rgb]{0,0,0}\makebox(0,0)[lt]{\lineheight{1.25}\smash{\begin{tabular}[t]{l}$\hat{n}$\end{tabular}}}}%
    \put(0,0){\includegraphics[width=\unitlength,page=2]{tube.pdf}}%
    \put(0.87584775,0.29822241){\makebox(0,0)[lt]{\lineheight{1.25}\smash{\begin{tabular}[t]{l}$\hat{B}$\end{tabular}}}}%
    \put(0,0){\includegraphics[width=\unitlength,page=3]{tube.pdf}}%
    \put(0.82116013,0.3537934){\makebox(0,0)[lt]{\lineheight{1.25}\smash{\begin{tabular}[t]{l}$n$\end{tabular}}}}%
    \put(0.85680707,0.36178754){\makebox(0,0)[lt]{\lineheight{1.25}\smash{\begin{tabular}[t]{l}$\beta$\end{tabular}}}}%
  \end{picture}%
\endgroup%

    \caption{The central magnetic line in black connects the two disks on the stars' respective surfaces.  
    A nearby magnetic line is the dashed green line with separation vector ${n}(s)$ and separation direction $\hat{n}(s)$. }
    \label{fig:tube}
\end{figure}

Consider a pair of disks\footnote{For global considerations $D_1$ is the magnetic North cap of star 1 and $D_2$ is the magnetic South cap of star 2.  For local considerations, these are small star surface disks centered on the beginning and end of some central magnetic line.} $D_1$ and $D_2$ (on the respective star surfaces)  each oriented by a unit normal $\vec{\mu}_1$ and $\vec{\mu}_2$ directed along the magnetic field as in Figure~\ref{fig:tube}. Their centers are connected by a length-parameterized path (representing a magnetic field line), {\em the central line},
\begin{align}
	\beta:[0,\ell]&\to\mathbb{R}^3\\
	s&\mapsto\beta(s) \nonumber
,\end{align}
with unit tangent $\hat{B}=\frac{d\beta}{ds}$. 
A nearby magnetic field line (dashed green) intersects a normal disk centered at $\beta(s)$ at some point  $\beta(s)+n(s)$ and thus the shape of the nearby line is determined by a normal vector field $n(s).$ Since we are only concerned with twisting, we keep track of the direction unit normal vector field $\hat{n}(s)=n(s)/|n(s)|.$ 
We shall be concerned with how much one line twists around the other, or, rather, how this twisting is changing with time, as the two discs orbit each other while also spinning at the same time.

For now let us focus on the static situation and describe the topology and geometry of a single magnetic tube, represented by a framed path $(\beta,\hat{n})$ with given initial and final frames specified by fixed  
$\hat{B}(0)=\vec{\mu}_1, \hat{B}(\ell)=\vec{\mu}_2,$ and fixed $\hat{n}(0)=\hat{n}_1$ and $\hat{n}(\ell)=\hat{n}_2$. 

\subsection{Paths in \texorpdfstring{$SO(3)$}{SO(3)}}
\label{sec:SO3}
The frame along a path $\beta$ is specified by a single normal field  $\hat{n}(s)$.  Of course the complete frame is formed by the triplet of vectors $f=(\hat{B},\hat{n},\hat{B}\times \hat{n}).$ Written as a $ 3 \times 3$ matrix, with each column representing a vector in this frame, we have an orthogonal matrix, thus a framed path gives a corresponding path in $SO(3)$ with
\begin{align}
	f: [0,\ell]&\to SO(3)\\
	s&\mapsto f(s)=(\hat{B}(s),\hat{n}(s),\hat{B}(s)\times \hat{n}(s))
.\end{align}
Thus, thinking of a small magnetic tube as a framed path we arrive at a path $f$ in $SO(3)$. Clearly, the first column of $f(s)$ is $\hat{B}(s)$, thus, the path $\beta$ can be recovered from  $f$ via integration $\beta(s) = \beta(0) + \int_0^s f(s) (1,0,0)^T ds$.

Given two such framed paths $(\beta,\hat{n})$ and $(\beta',\hat{n}')$ with the same initial and final frames, one might ask whether it is possible to deform one of them into another, while holding the initial and final frames fixed.  If it is indeed possible, then the corresponding paths $f$ and  $f'$ in $SO(3)$ are called homotopic.  
Thus it is worth understanding the space of paths in the group of orientation preserving rotations $SO(3).$ Any rotation of $\mathbb{R}^3$ can be specified by a single vector: its direction specified by the (oriented) axis of rotation and its length is the angle of the rotation around this axis.  
In particular the length of this vector does not exceed $\pi$.  At this stage all rotations appear to be in one-to-one correspondence with a radius $\pi$ ball in  $\mathbb{R}^3,$ except for one caveat. A rotation by angle $\pi$ around  $\hat{m}$ produces the same result as a rotation by $\pi$ around  $-\hat{m},$ therefore the antipodal points on the surface of this ball correspond to the same rotation.  One concludes that the group of rotations $SO(3)$ is a three-ball with antipodal points of its surface identified.  

One can view this three-ball as a Northern hemisphere of some three-sphere $S^3$, making clear the identification of $SO(3)$ with the three-dimensional projective space $ \mathbb{R}P^3=S^3/\{\pm 1\}$, which is a three-sphere with its antipodal points identified (or, equivalently, the space of lines in $\mathbb{R}^3$ passing through the origin): $\mathbb{R}P^3=SO(3)$.

While a three sphere is simply connected, i.e. any two paths between two given points can be deformed into each other. The projective space has exactly two classes of paths.  The difference between the two paths is a closed path obtained by traversing the first path and then returning backwards along the second path. This difference between the two paths in different classes is homotopic to (i.e. can be deformed to) the path in $S^3$ connecting two of its antipodal points. 

Note: the above is often illustrated with a belt, with its two ends held fixed at a distance, so that one can move the belt around these ends increasing or decreasing its twisting in the process.  However, if one breaks the rules and twists one of the ends once, the resulting belt configuration differs from all prior ones, in that those cannot be reached by manipulating the belt, while its ends are held fixed.  

The above correspondence is often interpreted in terms of spin.  Namely, identifying $SU(2)$ as a three-sphere (for example, as a sphere of unit quaternions), and thinking of the three space $\mathbb{R}^3$ as imaginary quaternions, the rotation $R(M)$ of $\mathbb{R}^3$ is a conjugation of the imaginary quaternions by a unit quaternion: $SU(2)\ni M: x\mapsto M^{-1}x M$. Then, clearly, $M$ and  $-M$ have exactly the same effect on $\mathbb{R}^3$ and the group of all orientation preserving rotations  $SO(3)=SU(2)/\{\pm 1\}$. A framed path gives a path in $SO(3)$ (for example starting at identity and arriving at some fixed frame $M\in SO(3)$) and lifts unambiguously to a path in  $SU(2)$ (also beginning at identity). It will arrive at some element of $\hat{M}\in SU(2)$. 
As we deform the path in $SO(3)$, its lift to $SU(2)$ will still be beginning at identity and ending at  $\hat{M}.$  However, there is a whole other class of paths in $SO(3)$ to  $M$, they lift to paths in  $SU(2)$ from identity to  $-\hat{M}$.  Thus what governs the framed path  (up to its deformations) is the spin group $Spin(3)=SU(2),$ rather than $SO(3).$
This is the reason a belt is sometimes used to illustrating fermion statistics.

\subsection{Paths in the unit tangent bundle of a two-sphere}
\label{sec:F}
\begin{figure}[!ht]
    \centering
	\def\svgwidth{1\columnwidth}
\begingroup%
  \makeatletter%
  \providecommand\color[2][]{%
    \errmessage{(Inkscape) Color is used for the text in Inkscape, but the package 'color.sty' is not loaded}%
    \renewcommand\color[2][]{}%
  }%
  \providecommand\transparent[1]{%
    \errmessage{(Inkscape) Transparency is used (non-zero) for the text in Inkscape, but the package 'transparent.sty' is not loaded}%
    \renewcommand\transparent[1]{}%
  }%
  \providecommand\rotatebox[2]{#2}%
  \newcommand*\fsize{\dimexpr\f@size pt\relax}%
  \newcommand*\lineheight[1]{\fontsize{\fsize}{#1\fsize}\selectfont}%
  \ifx\svgwidth\undefined%
    \setlength{\unitlength}{680.31496063bp}%
    \ifx\svgscale\undefined%
      \relax%
    \else%
      \setlength{\unitlength}{\unitlength * \real{\svgscale}}%
    \fi%
  \else%
    \setlength{\unitlength}{\svgwidth}%
  \fi%
  \global\let\svgwidth\undefined%
  \global\let\svgscale\undefined%
  \makeatother%
  \begin{picture}(1,0.5)%
    \lineheight{1}%
    \setlength\tabcolsep{0pt}%
    \put(0,0){\includegraphics[width=\unitlength,page=1]{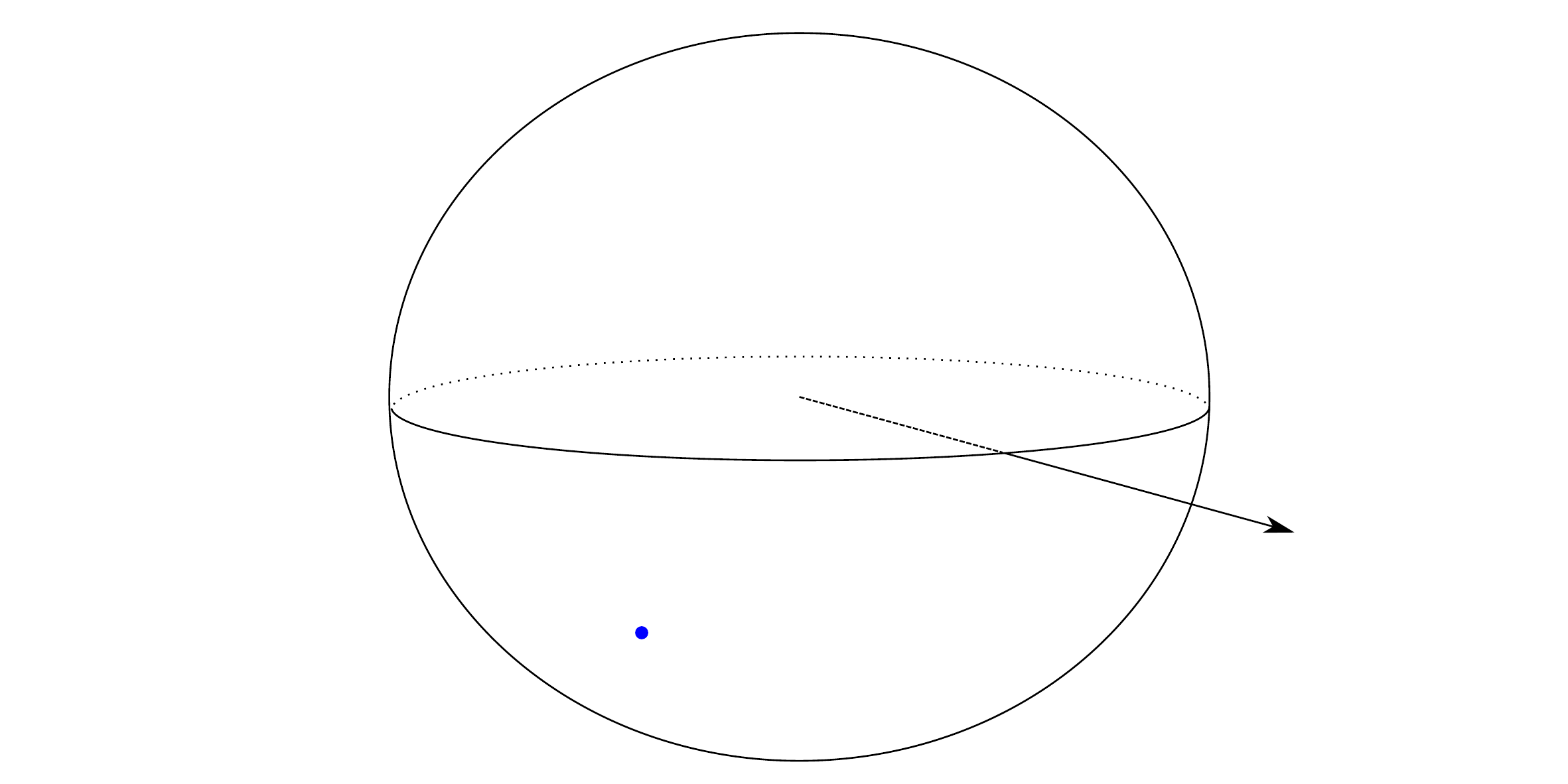}}%
    \put(0.8125569,0.1715196){\color[rgb]{0,0,0}\makebox(0,0)[lt]{\lineheight{1.25}\smash{\begin{tabular}[t]{l}$\vec{R}$\end{tabular}}}}%
    \put(0.35122836,0.34258416){\color[rgb]{0,0,0}\makebox(0,0)[lt]{\lineheight{1.25}\smash{\begin{tabular}[t]{l}$\vec{\mu}_1$\end{tabular}}}}%
    \put(0.40400362,0.11510468){\color[rgb]{0,0,0}\makebox(0,0)[lt]{\lineheight{1.25}\smash{\begin{tabular}[t]{l}$\vec{\mu}_2$\end{tabular}}}}%
    \put(0,0){\includegraphics[width=\unitlength,page=2]{dirs.pdf}}%
  \end{picture}%
\endgroup%

    \caption{A single magnetic line (with displacement $\vec{R}$) is represented by a blue path in the sphere of directions (satisfying $\int_0^\ell \hat{B}(s)ds=\vec{R}$).}
    \label{fig:dirs}
\end{figure}

\begin{figure}[!ht]
    \centering
	\def\svgwidth{1\columnwidth}
\begingroup%
  \makeatletter%
  \providecommand\color[2][]{%
    \errmessage{(Inkscape) Color is used for the text in Inkscape, but the package 'color.sty' is not loaded}%
    \renewcommand\color[2][]{}%
  }%
  \providecommand\transparent[1]{%
    \errmessage{(Inkscape) Transparency is used (non-zero) for the text in Inkscape, but the package 'transparent.sty' is not loaded}%
    \renewcommand\transparent[1]{}%
  }%
  \providecommand\rotatebox[2]{#2}%
  \newcommand*\fsize{\dimexpr\f@size pt\relax}%
  \newcommand*\lineheight[1]{\fontsize{\fsize}{#1\fsize}\selectfont}%
  \ifx\svgwidth\undefined%
    \setlength{\unitlength}{680.31496063bp}%
    \ifx\svgscale\undefined%
      \relax%
    \else%
      \setlength{\unitlength}{\unitlength * \real{\svgscale}}%
    \fi%
  \else%
    \setlength{\unitlength}{\svgwidth}%
  \fi%
  \global\let\svgwidth\undefined%
  \global\let\svgscale\undefined%
  \makeatother%
  \begin{picture}(1,0.5)%
    \lineheight{1}%
    \setlength\tabcolsep{0pt}%
    \put(0,0){\includegraphics[width=\unitlength,page=1]{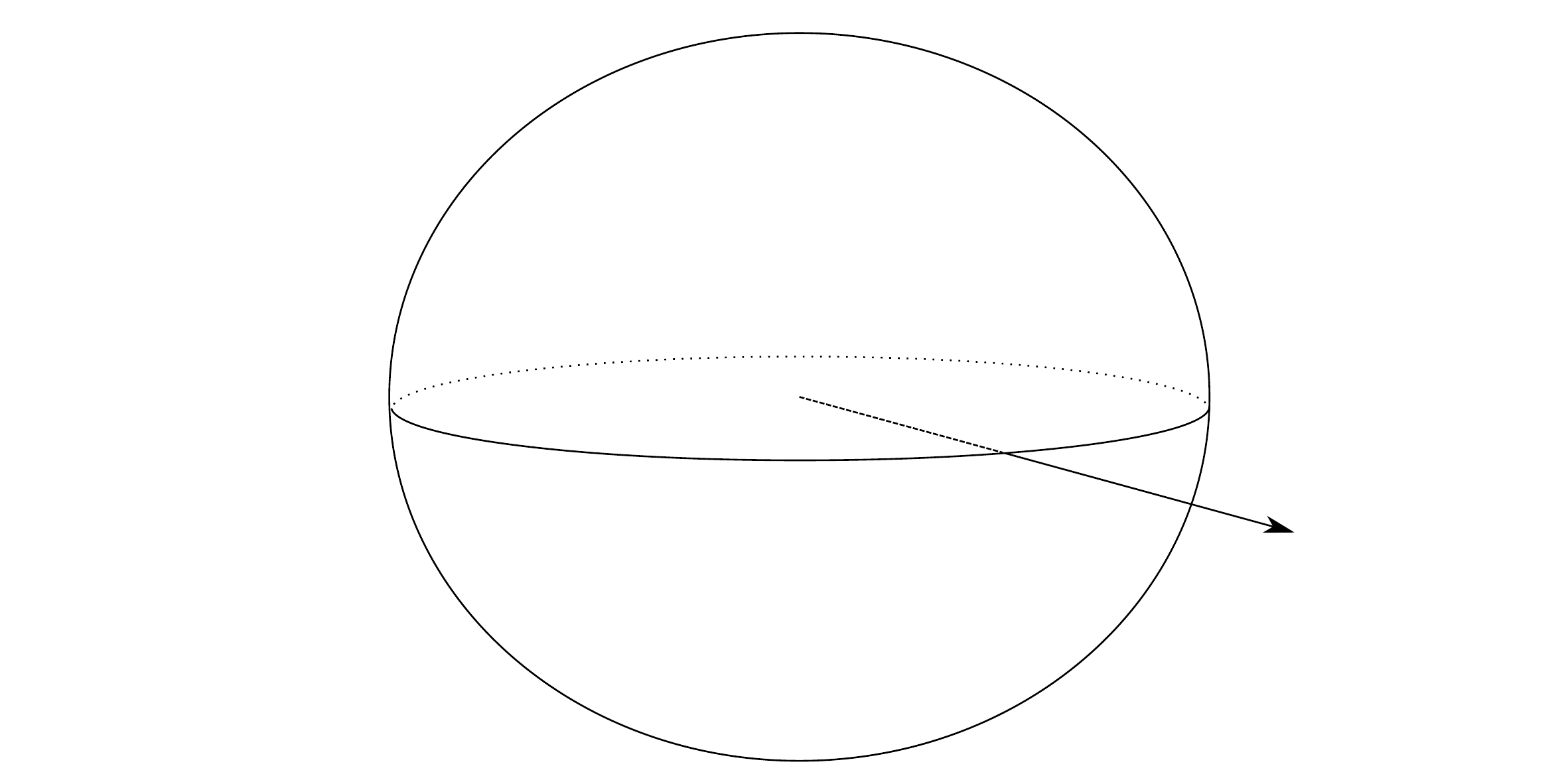}}%
    \put(0.8125569,0.1715196){\color[rgb]{0,0,0}\makebox(0,0)[lt]{\lineheight{1.25}\smash{\begin{tabular}[t]{l}$\vec{R}$\end{tabular}}}}%
    \put(0.35122836,0.34258416){\color[rgb]{0,0,0}\makebox(0,0)[lt]{\lineheight{1.25}\smash{\begin{tabular}[t]{l}$\vec{\mu}_1$\end{tabular}}}}%
    \put(0.40400362,0.11510468){\color[rgb]{0,0,0}\makebox(0,0)[lt]{\lineheight{1.25}\smash{\begin{tabular}[t]{l}$\vec{\mu}_2$\end{tabular}}}}%
    \put(0,0){\includegraphics[width=\unitlength,page=2]{dirs2.pdf}}%
  \end{picture}%
\endgroup%

    \caption{A magnetic tube is represented by the green {\bf dashed} path in $\mathbb{F}$, which is equivalent to a unit tangent vector field along the blue {\bf solid} line.}
    \label{fig:dirs2}
\end{figure}

Perhaps, a more economical view of a framed path is in terms of the tangent bundle $TS^2$ of a two-sphere of directions $S^2.$ This bundle consists of disjoint tangent planes to $S^{2}$. Namely, for any length parameterized  path $\beta$ in  $\mathbb{R}^3$, its normal tangents $\hat{B}:[0,\ell]\to S^2, s\mapsto \hat{B}(s)=\frac{d}{ds}\beta(s)$ form a path in the unit sphere $S^2$ of directions, as in Fig.\ref{fig:dirs}, while $\hat{n}(s)$, orthogonal unit vector to $\hat{B}(s)$, is a tangent vector to $S^2$ at  $\hat{B}(s)$ as in Fig.~\ref{fig:dirs2}. Since it is normalized, it lies in the unit circle in the tangent plane $T_{\hat{B}(s)}S^2$ to $S^2$ at point  $\hat{B}(s).$ 

Thus we arrive at a slightly different picture of the space $\mathbb{F}$ of all unit circles in the tangent spaces of the unit two-sphere: $\mathbb{F}=T^1S^2\subset TS^2.$ An element of this space is specified by a pair $(a,b)$ of orthogonal unit vectors, the first vector $a$ specifies the point in  $S^2$ and the second vector  $b$ specifies a point in the tangent plane to  $S^2$ at  $a.$ 
Of course, given a point $(a,b)\in \mathbb{F}$ the forgetful map $(a,b)\mapsto a$ gives the corresponding element of the sphere, 
in other words we have a natural projection $\mathbb{F}\to S^2, (a,b)\mapsto a$ (it projects a circle fiber of normals to its corresponding point on the sphere of directions).

A framed path $(\beta,\hat{n})$ gives a path $\hat{\beta}$ in $\mathbb{F}$ by
\begin{align}
	\hat{\beta}: [0,\ell]&\to \mathbb{F}\\
	s&\mapsto (\hat{B}(s),\hat{n}(s))\nonumber
.\end{align}
The path in the sphere traced by $\hat{B}(s)$ simply traces out the directions of the path $\beta,$ while the unit tangent field  $\hat{n}(s)$ along this path lifts it to $\mathbb{F}.$ The path in the sphere begins at $\vec{\mu}_1$ and ends in $\vec{\mu}_2.$ It is constrained by $\vec{R}=\int_0^{\ell}\hat{B}(s) ds.$ Thus, this path is bound to spend some time near $\vec{R}/|\vec{R}|\in S^2,$ as illustrated in Figure~\ref{fig:dirs}.  
Needless to say, our description is equivalent to that of section~\ref{sec:SO3}, as the space $\mathbb{F}$ can be identified with $SO(3)$ via  $(a,b)\mapsto(a,b,a\times b).$

Let us comment on the topology of the unit circle bundle $T^1S^2$ of the two-sphere $S^2$. In particular, we would like to make it transparent that it is a degree 2 Hopf bundle. This degree 2 is at the heart of the effect discussed in this paper;  it is responsible for the persistence of the 2:1 resonance. 
Perhaps the most common way to demonstrate that the degree is 2 is to consider the polar projection $F:S \setminus \{N\}\to \mathbb{R}^2 $ from the North pole $N$ to the plane tangent at the South pole  $S$.  Taking a constant vector field on the plane and pulling it back to $S\setminus \{N\} $ one obtains a vector field that winds twice as one circumnavigates the North pole traversing a small meridian near it. 
\begin{figure}[!ht]
\centering
\includegraphics[width=.7\textwidth]{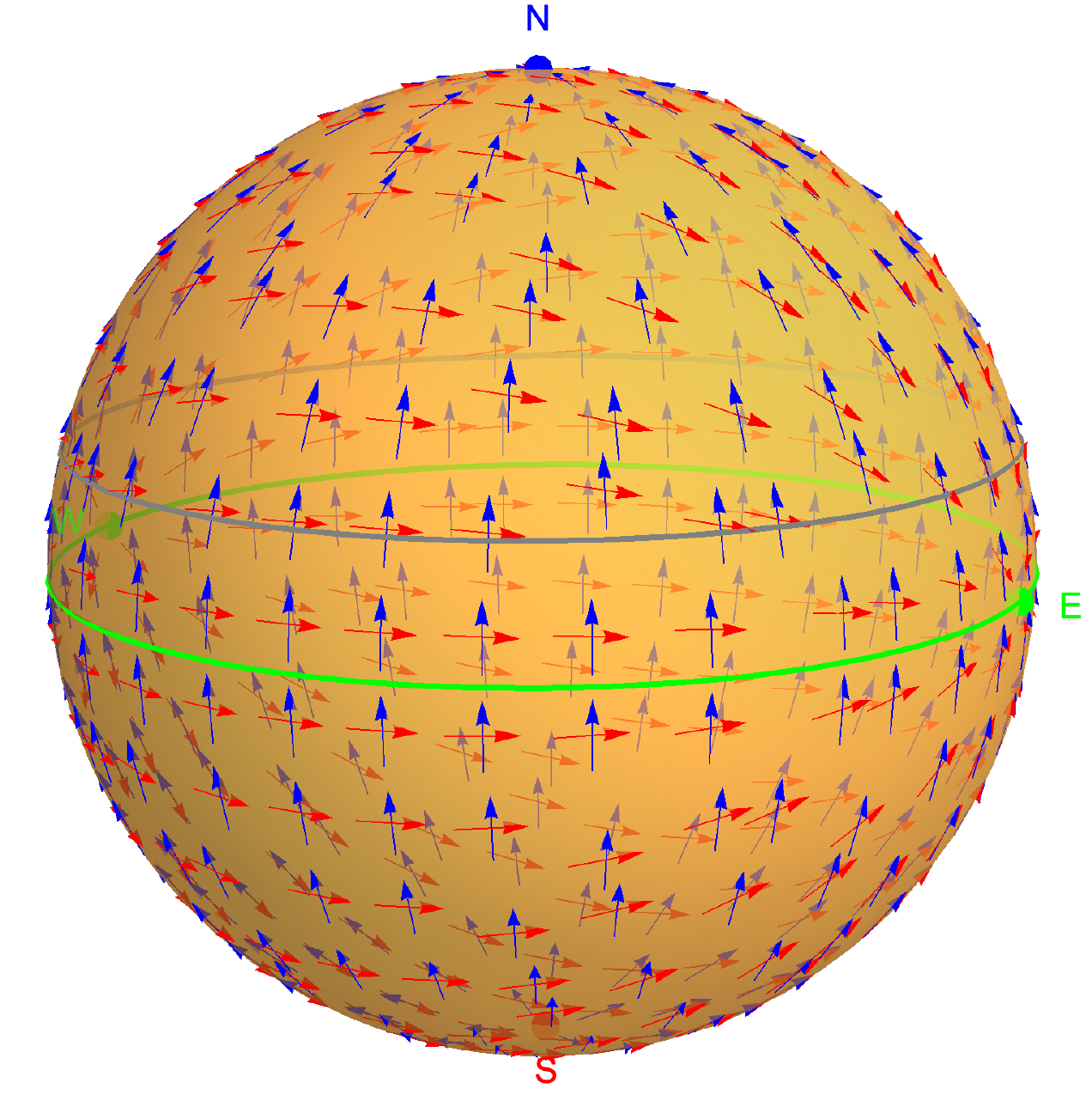}
\caption{The vector field $v$ is in blue is due North and the field  $u$ in red is directed towards the East pole. As one circles Eastwards along the gray parallel, $v$ rotates once counterclockwise relative to  $u$.}
\label{T1S2} 
\end{figure}

For our purposes, however, we use a different picture. We trivialize the bundle over the Northern hemisphere  $U_N,$ the southern hemisphere  $U_S,$ and over the {\em tropical region} $U_T=S^2\setminus \{N,S\} $ containing the equator, as follows.  
Let us mark a point $E$ on the equator and call it the {\em East pole}.  We also call its antipodal point the {\em West pole}.
A unit tangent vector field $u$ directed towards $E$ is then well defined everywhere on the sphere except at  $E$ and  at $W$. We use it to define coordinates in the tangent circle fiber over  $U_N$ and  $U_S.$  For any point  $p\in U_N$ and for any tangent unit vector $w_p$ at  $p$, let $\phi_N(w_p)$ be the angle between $u_p$ and $w_p$.  
Now, any real number determines a unique unit tangent vector at  $p$, whose angle with $u_p$ is that number. In other words we defined a coordinate along the circle fiber. 
Similarly, if $p\in U_S,$ we define $\phi_S(w_p)$ to be the angle between $u_p$ and  $w_p.$ 

Next, consider the Northward unit tangent field $v$.  It is well defined everywhere in the tropical region  $U_T$. For any $p\in U_T$ let $\phi_T(w_p)$ be the angle between $v_p$ and  $w_p.$  

Consider some parallel in the Northern hemisphere and a point $p$ on it. Clearly,  $\phi_T = \phi_N +\phi_T(v_p),$ i.e. the difference between the two angles $\phi_T(w_p)$ and  $\phi_N(w_p)$ is the angle between  $v_p$ and $u_p$. Whenever the point $p$ circumnavigates the parallel Eastward, this angle between  $v_p$ and  $u_p$ changes by  $2\pi$, i.e. $\Delta(\phi_T(w_p)-\phi_N(w_p))=2\pi.$
Similarly, circumnavigating any parallel in the Southern hemisphere Eastwards, the change in the difference between $\phi_T$ and  $\phi_S$ is  $-2\pi,$ i.e. 
$\Delta(\phi_T(w_p)-\phi_S(w_p))=-2\pi.$ 

For example, if we have a path in  $\mathbb{F}$ passing over the equator, then, if we move it all the way around the equator holding the limiting value of, say, $\phi_N$ at the equator fixed, then the value of  $\phi_E$ will change by $2\pi,$ while the value of $\phi_S$ will change by  $4\pi.$ 
It is this factor of 2 in  $2\times 2\pi$, that is the {\em degree} or the {\em Hopf number} of this bundle. (It is exactly because this degree is not zero that the sphere `cannot be combed', i.e. it has no nowhere vanishing continuous tangent fields.  It is also the reason there is no global coordinate along the fiber of $\mathbb{F}$.)

\subsection{Relative twisting}
For some paths one might be able to introduce the notion of twisting, for example, using the (unit) tangent $\hat{B}$ and curvature $d\hat{B}/ds$ of the path to form a frame along it. Such a Frenet frame could be used to define the twisting of a magnetic tube centered around that path. However, as we consider the evolution of the path, as it changes its shape we might (and even are likely to) loose this frame. For example, the curvature might vanish at some point.  This would prevent continuous tracking of this path twisting angle.  
In other words, there are two issues preventing us from discussing twisting of an arbitrary tube: 
\begin{enumerate}
	\item 
there cannot be a coordinate on the fiber circle that is valid over the whole sphere of directions and 
\item
there is no guarantee that the path of directions $\hat{B}$ is differentiable, even though the original path in $\mathbb{R}^3$ was. 
\end{enumerate}
However, if we have one path  $\hat{\gamma}_1$ and another path $\hat{\gamma}_2$ in $T^1S^2,$ and both happen to project to the same path $\gamma$ in $S^2,$ then there is a good notion of  how much $\hat{\gamma}_2$ winds around $\hat{\gamma}_1.$ 
To do this, as we traverse $\gamma$, we continuously track the change in the angle between $\hat{\gamma}_1(s)$ and $\hat{\gamma}_2(s)$. Importantly, this does not involve any charts, fiber coordinates, or auxiliary vector fields.
Thus, for such pair of paths the relative twisting angle is well defined.

In general we shall consider a family of framed paths $\hat{\gamma}(s,t)$, where $s$ is the length parameter along a path at time  $t$.  In particular,  $\hat{\gamma}(s,0)=\hat{\gamma}_1(s)$ is the initial path and $\hat{\gamma}(s,2\pi)=\hat{\gamma}_2(s)$ is the final path.
Also, the projection of $\hat{\gamma}(s,t)\in\mathbb{F}$ is the direction $\gamma(s,t).$ 

Let us begin with a very simple example, consider a path $\gamma(s,t)$ that is the meridian from  $N$ to  $S$ with longitude  $t$. Clearly, as  $t$ changes from  $0$ to  $2\pi$ this meridian sweeps the sphere and returns to its original shape, i.e.  $\gamma(s,0)=\gamma(s,2\pi).$ 
Consider a tangent vector field to each meridian that, away from a small neighborhood of the equator is given by the East pole directed field  $v$. 
(Clearly, the field $v$ does not give a good normal field along the whole family $\gamma(s,t)$, since it is not defined at $E$ and $W$. Thus, for each $t$ we interpolate the resulting tangent field along the path  $\gamma(s,t)$ in the vicinity of the equator.)  
Let $\hat{\gamma}(s,0)$ be the path lifting  the meridian $\gamma(s,0)$ using that field (outside the neighborhood of the equator) and continuous along the whole meridian. As we vary  $t$ and the meridian  $\gamma(s,t)$ rotates sweeping the sphere, we keep  $\hat{\gamma}(s,t)$ given by $v$ away from the equator and change continuously with  $s$ and  $t$. We would like to learn about the resulting path  $\hat{\gamma}(s,2\pi)$ relation to the original path $\hat{\gamma}(s,0).$ 

\begin{figure}[!ht]
    \centering
	\def\svgwidth{0.7\columnwidth}
\begingroup%
  \makeatletter%
  \providecommand\color[2][]{%
    \errmessage{(Inkscape) Color is used for the text in Inkscape, but the package 'color.sty' is not loaded}%
    \renewcommand\color[2][]{}%
  }%
  \providecommand\transparent[1]{%
    \errmessage{(Inkscape) Transparency is used (non-zero) for the text in Inkscape, but the package 'transparent.sty' is not loaded}%
    \renewcommand\transparent[1]{}%
  }%
  \providecommand\rotatebox[2]{#2}%
  \newcommand*\fsize{\dimexpr\f@size pt\relax}%
  \newcommand*\lineheight[1]{\fontsize{\fsize}{#1\fsize}\selectfont}%
  \ifx\svgwidth\undefined%
    \setlength{\unitlength}{402.12339867bp}%
    \ifx\svgscale\undefined%
      \relax%
    \else%
      \setlength{\unitlength}{\unitlength * \real{\svgscale}}%
    \fi%
  \else%
    \setlength{\unitlength}{\svgwidth}%
  \fi%
  \global\let\svgwidth\undefined%
  \global\let\svgscale\undefined%
  \makeatother%
  \begin{picture}(1,0.7790378)%
    \lineheight{1}%
    \setlength\tabcolsep{0pt}%
    \put(0,0){\includegraphics[width=\unitlength,page=1]{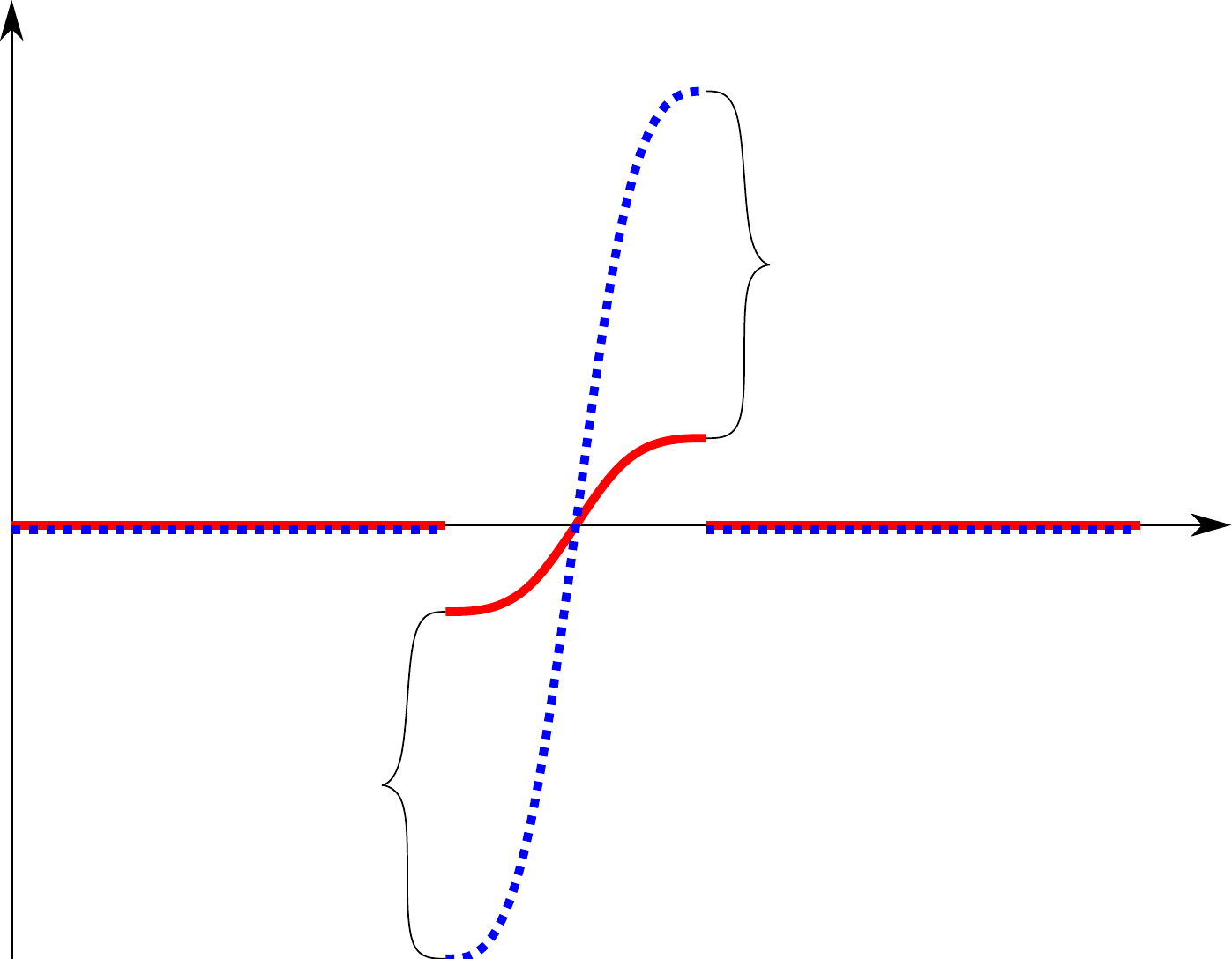}}%
    \put(0.64391377,0.56393547){\makebox(0,0)[lt]{\lineheight{1.25}\smash{\begin{tabular}[t]{l}$2\pi$\end{tabular}}}}%
    \put(0.23127037,0.12878415){\makebox(0,0)[lt]{\lineheight{1.25}\smash{\begin{tabular}[t]{l}$2\pi$\end{tabular}}}}%
    \put(0.09044935,0.37981164){\makebox(0,0)[lt]{\lineheight{1.25}\smash{\begin{tabular}[t]{l}$\phi_N(s)$\end{tabular}}}}%
    \put(0.66064759,0.37418464){\makebox(0,0)[lt]{\lineheight{1.25}\smash{\begin{tabular}[t]{l}$\phi_S(s)$\end{tabular}}}}%
    \put(0.42837668,0.0687634){\makebox(0,0)[lt]{\lineheight{1.25}\smash{\begin{tabular}[t]{l}$\phi_T(s,2\pi)$\end{tabular}}}}%
    \put(0.31583757,0.30025136){\makebox(0,0)[lt]{\lineheight{1.25}\smash{\begin{tabular}[t]{l}$\phi_T(s,0)$\end{tabular}}}}%
  \end{picture}%
\endgroup%

    \caption{Local fiber coordinate of the initial path in solid red and of the final path in dashed blue.  North and South patch coordinates $\phi_N(s,t)$ and  $\phi_S(s,t)$ are held fixed and equal to zero.}
    \label{fig:graph}
\end{figure}
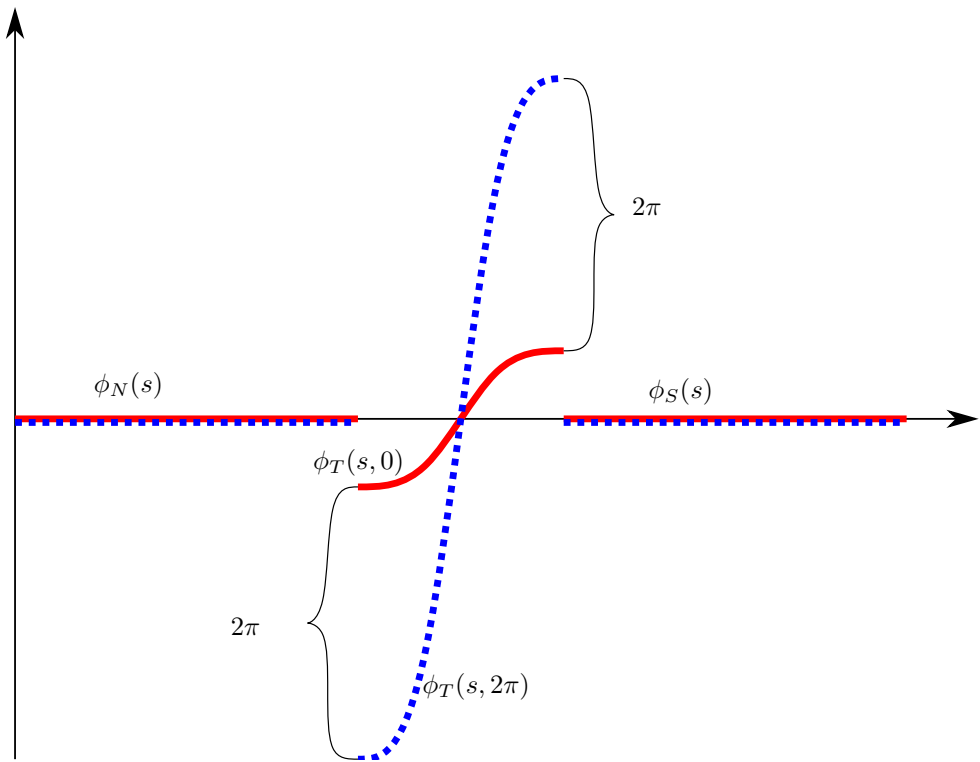

In terms of the local coordinates, these paths have $\phi_N=0$ and $\phi_S=0$ away from the neighborhood of the equator. There is a discontinuity between $\phi_N$ and  $\phi_T$, for example, and this discontinuity changes by  $2\pi$ as the meridian sweeps the sphere. 
Since we keep the North and South coordinates zero by construction, and the gap between $\phi_N$ and  $\phi_T$ changes by  $2\pi$, the interpolating path near equator has to adjust, changing its initial value of  $\phi_T$ just North of equator.  Similarly, since the gap between  $\phi_T$ and  $\phi_S$ changes by  $2\pi$, the value of  $\phi_T$ just South of the equator changes.  As a result, the interpolating part of the path shown in dashed blue in Fig.~\ref{fig:graph} wraps twice around the fiber relative to the original path shown in solid red.

In fact, nothing in this argument depends on our special choice of path lifting, so long as we hold the initial and final frames fixed.
(The {\bf solid} red and {\bf dashed} blue graphs in Fig.~\ref{fig:graph} will be arbitrary, but the relation between their discontinuities will remain the same, adding up to $4\pi$.) 
As the path $\gamma$ sweeps the sphere once (moving Eastwards), the path  $\hat{\gamma}(s,2\pi)$ will twist by $4\pi$ relative to  $\hat{\gamma}(s,0).$
\subsection[]{Case redux}
\label{sec:case_redux}
With our newly gained perspective let us revisit the cases discussed in Sec.~\ref{sec:cases} and account for the observed twisting.

\begin{figure}[ht]
    \centering
	\begin{tabular}[c]{cc}
	\def\svgwidth{0.45\columnwidth}
\begingroup%
  \makeatletter%
  \providecommand\color[2][]{%
    \errmessage{(Inkscape) Color is used for the text in Inkscape, but the package 'color.sty' is not loaded}%
    \renewcommand\color[2][]{}%
  }%
  \providecommand\transparent[1]{%
    \errmessage{(Inkscape) Transparency is used (non-zero) for the text in Inkscape, but the package 'transparent.sty' is not loaded}%
    \renewcommand\transparent[1]{}%
  }%
  \providecommand\rotatebox[2]{#2}%
  \newcommand*\fsize{\dimexpr\f@size pt\relax}%
  \newcommand*\lineheight[1]{\fontsize{\fsize}{#1\fsize}\selectfont}%
  \ifx\svgwidth\undefined%
    \setlength{\unitlength}{453.54330709bp}%
    \ifx\svgscale\undefined%
      \relax%
    \else%
      \setlength{\unitlength}{\unitlength * \real{\svgscale}}%
    \fi%
  \else%
    \setlength{\unitlength}{\svgwidth}%
  \fi%
  \global\let\svgwidth\undefined%
  \global\let\svgscale\undefined%
  \makeatother%
  \begin{picture}(1,0.75)%
    \lineheight{1}%
    \setlength\tabcolsep{0pt}%
    \put(0,0){\includegraphics[width=\unitlength,page=1]{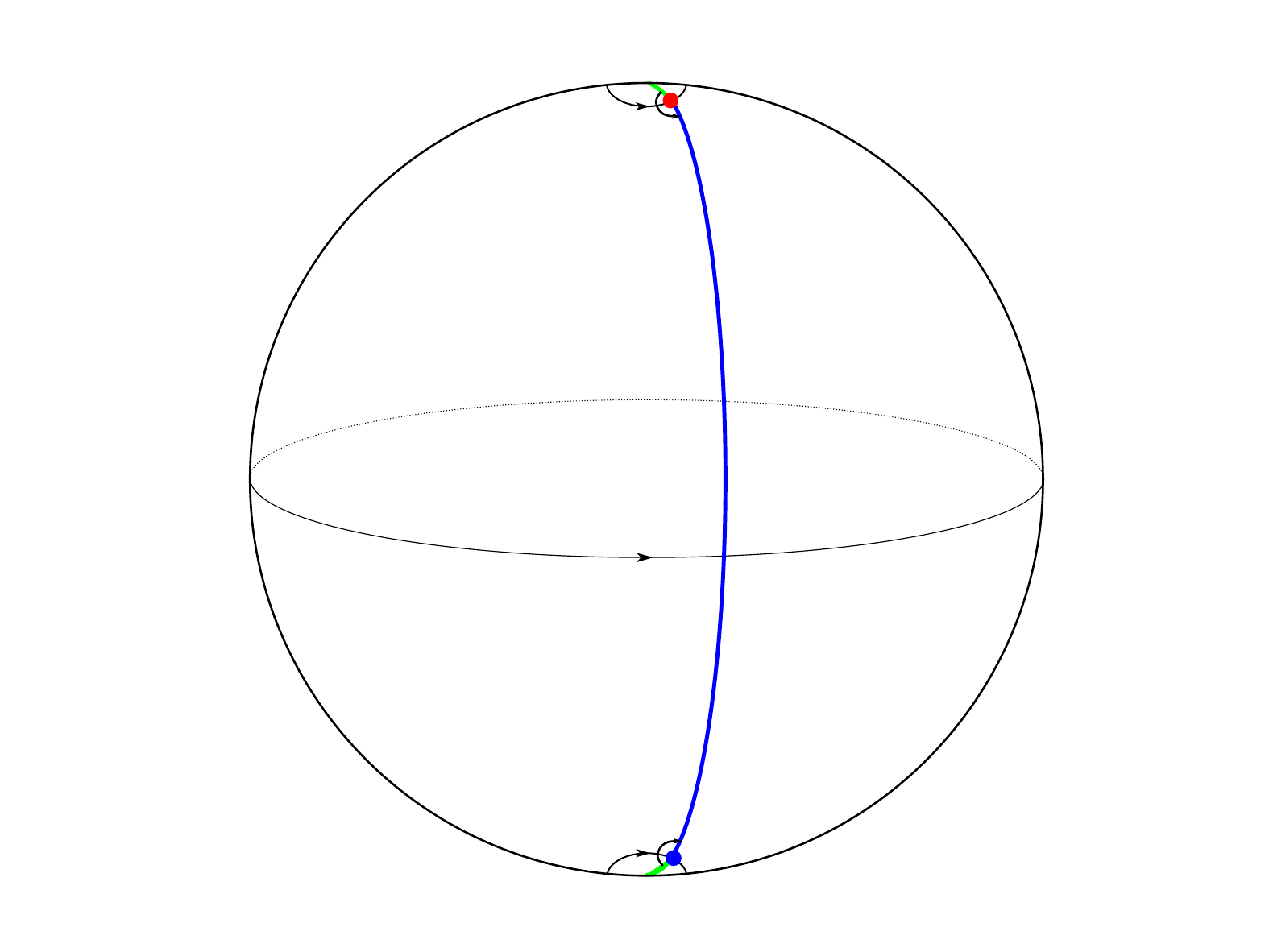}}%
    \put(0.509045,0.70670362){\makebox(0,0)[lt]{\lineheight{1.25}\smash{\begin{tabular}[t]{l}$\vec{\mu}_1$\end{tabular}}}}%
    \put(0.50424229,0.01717249){\makebox(0,0)[lt]{\lineheight{1.25}\smash{\begin{tabular}[t]{l}$\vec{\mu}_2$\end{tabular}}}}%
    \put(0.5961887,0.27230173){\makebox(0,0)[lt]{\lineheight{1.25}\smash{\begin{tabular}[t]{l}$\hat{R}$\end{tabular}}}}%
    \put(0.48153998,0.62832485){\makebox(0,0)[lt]{\lineheight{1.25}\smash{\begin{tabular}[t]{l}$\chi_1$\end{tabular}}}}%
    \put(0.48038489,0.10127245){\makebox(0,0)[lt]{\lineheight{1.25}\smash{\begin{tabular}[t]{l}$\chi_2$\end{tabular}}}}%
    \put(0,0){\includegraphics[width=\unitlength,page=2]{case1.pdf}}%
    \put(0.29821869,0.55148283){\makebox(0,0)[lt]{\lineheight{1.25}\smash{\begin{tabular}[t]{l}$\vec{\omega}_1=\vec{\omega}_2=\vec{\Omega}$\end{tabular}}}}%
    \put(0,0){\includegraphics[width=\unitlength,page=3]{case1.pdf}}%
  \end{picture}%
\endgroup%

    &
	\def\svgwidth{0.45\columnwidth}
\begingroup%
  \makeatletter%
  \providecommand\color[2][]{%
    \errmessage{(Inkscape) Color is used for the text in Inkscape, but the package 'color.sty' is not loaded}%
    \renewcommand\color[2][]{}%
  }%
  \providecommand\transparent[1]{%
    \errmessage{(Inkscape) Transparency is used (non-zero) for the text in Inkscape, but the package 'transparent.sty' is not loaded}%
    \renewcommand\transparent[1]{}%
  }%
  \providecommand\rotatebox[2]{#2}%
  \newcommand*\fsize{\dimexpr\f@size pt\relax}%
  \newcommand*\lineheight[1]{\fontsize{\fsize}{#1\fsize}\selectfont}%
  \ifx\svgwidth\undefined%
    \setlength{\unitlength}{453.54330709bp}%
    \ifx\svgscale\undefined%
      \relax%
    \else%
      \setlength{\unitlength}{\unitlength * \real{\svgscale}}%
    \fi%
  \else%
    \setlength{\unitlength}{\svgwidth}%
  \fi%
  \global\let\svgwidth\undefined%
  \global\let\svgscale\undefined%
  \makeatother%
  \begin{picture}(1,0.75)%
    \lineheight{1}%
    \setlength\tabcolsep{0pt}%
    \put(0,0){\includegraphics[width=\unitlength,page=1]{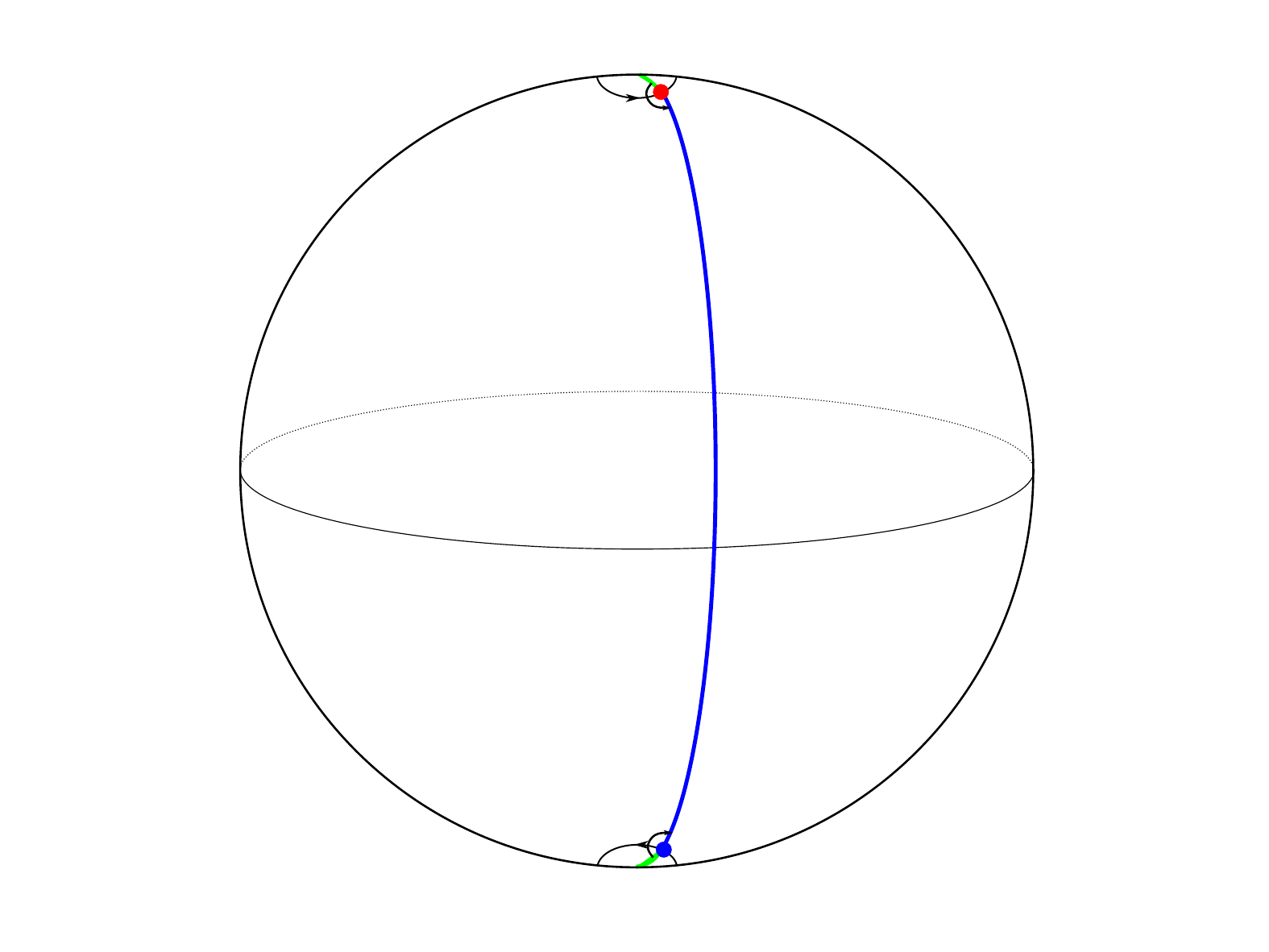}}%
    \put(0.48309944,0.7100112){\makebox(0,0)[lt]{\lineheight{1.25}\smash{\begin{tabular}[t]{l}$\vec{\mu}_1$\end{tabular}}}}%
    \put(0.48309944,0.02709409){\makebox(0,0)[lt]{\lineheight{1.25}\smash{\begin{tabular}[t]{l}$\vec{\mu}_2$\end{tabular}}}}%
    \put(0.58850964,0.27891637){\makebox(0,0)[lt]{\lineheight{1.25}\smash{\begin{tabular}[t]{l}$\hat{R}$\end{tabular}}}}%
    \put(0.47386091,0.63493968){\makebox(0,0)[lt]{\lineheight{1.25}\smash{\begin{tabular}[t]{l}$\chi_1$\end{tabular}}}}%
    \put(0.47270582,0.1078867){\makebox(0,0)[lt]{\lineheight{1.25}\smash{\begin{tabular}[t]{l}$\chi_2$\end{tabular}}}}%
    \put(0,0){\includegraphics[width=\unitlength,page=2]{case2.pdf}}%
    \put(0.43934115,0.56618396){\makebox(0,0)[lt]{\lineheight{1.25}\smash{\begin{tabular}[t]{l}$\vec{\omega}_1$\end{tabular}}}}%
    \put(0.3016609,0.20057946){\makebox(0,0)[lt]{\lineheight{1.25}\smash{\begin{tabular}[t]{l}$\vec{\omega}_2=-\vec{\omega}_1$\end{tabular}}}}%
    \put(0,0){\includegraphics[width=\unitlength,page=3]{case2.pdf}}%
  \end{picture}%
\endgroup%

    \end{tabular}
    \caption{Cases I and II in terms of the direction two-sphere. Left corresponds to Case I and right corresponds to Case II illustrated in Fig.~\ref{twisted} (rotation of this right figure illustrates Fig.~\ref{New-Final}). All have no winding of magnetic tubes.}
    \label{fig:case12}
\end{figure}

Case I had $\vec{\omega}_1=\vec{\omega}_2=\vec{\Omega}\sim \hat{z}$. In terms of the sphere of directions this can be visualized, left Fig.~\ref{fig:case12}, as points $\vec{\mu}_1$ and $\vec{\mu}_2$ presented as, respectively, red and blue dots circling in the same (Eastward) direction around, respectively, North and South poles.  The path of directions is the blue path on the sphere connecting these two points and passing in the vicinity of the black dot  $\hat{R}$. The latter rotates around the equator.  The twisting is captured by the sum of the angles $\chi_1$ and  $\chi_2.$ For Case I all three points (red, blue, and black) rotate with the angular velocity, and the resulting angles $\chi_1$ and  $\chi_2$ can remain constant.  Therefore, we observe no winding.

Case II, as illustrated in Fig.~\ref{twisted}, is very similar to the picture above. As illustrated in the right Fig.~\ref{fig:case12}, the red and blue dots moving in the opposite directions (with equal angular velocity), while the black bot is staying put.  In this case one of the $\chi$ angles increases, while another decreases, with their sum remaining constant (on average).  Therefore, there cannot be twisting in this case.

Case II, as illustrated in Fig.~\ref{New-Final}, is the picture we just described, but rotated, so that now the blue dot is staying put, the black dot rotates, and the red dot rotates twice as fast.  This modification has no bearing on the resulting evolution of the angles  $\chi_1$ and  $\chi_2$, again resulting in no winding.

\begin{figure}[ht]
    \centering
	\begin{tabular}[c]{cc}
	\def\svgwidth{0.45\columnwidth}
\begingroup%
  \makeatletter%
  \providecommand\color[2][]{%
    \errmessage{(Inkscape) Color is used for the text in Inkscape, but the package 'color.sty' is not loaded}%
    \renewcommand\color[2][]{}%
  }%
  \providecommand\transparent[1]{%
    \errmessage{(Inkscape) Transparency is used (non-zero) for the text in Inkscape, but the package 'transparent.sty' is not loaded}%
    \renewcommand\transparent[1]{}%
  }%
  \providecommand\rotatebox[2]{#2}%
  \newcommand*\fsize{\dimexpr\f@size pt\relax}%
  \newcommand*\lineheight[1]{\fontsize{\fsize}{#1\fsize}\selectfont}%
  \ifx\svgwidth\undefined%
    \setlength{\unitlength}{453.54330709bp}%
    \ifx\svgscale\undefined%
      \relax%
    \else%
      \setlength{\unitlength}{\unitlength * \real{\svgscale}}%
    \fi%
  \else%
    \setlength{\unitlength}{\svgwidth}%
  \fi%
  \global\let\svgwidth\undefined%
  \global\let\svgscale\undefined%
  \makeatother%
  \begin{picture}(1,0.75)%
    \lineheight{1}%
    \setlength\tabcolsep{0pt}%
    \put(0,0){\includegraphics[width=\unitlength,page=1]{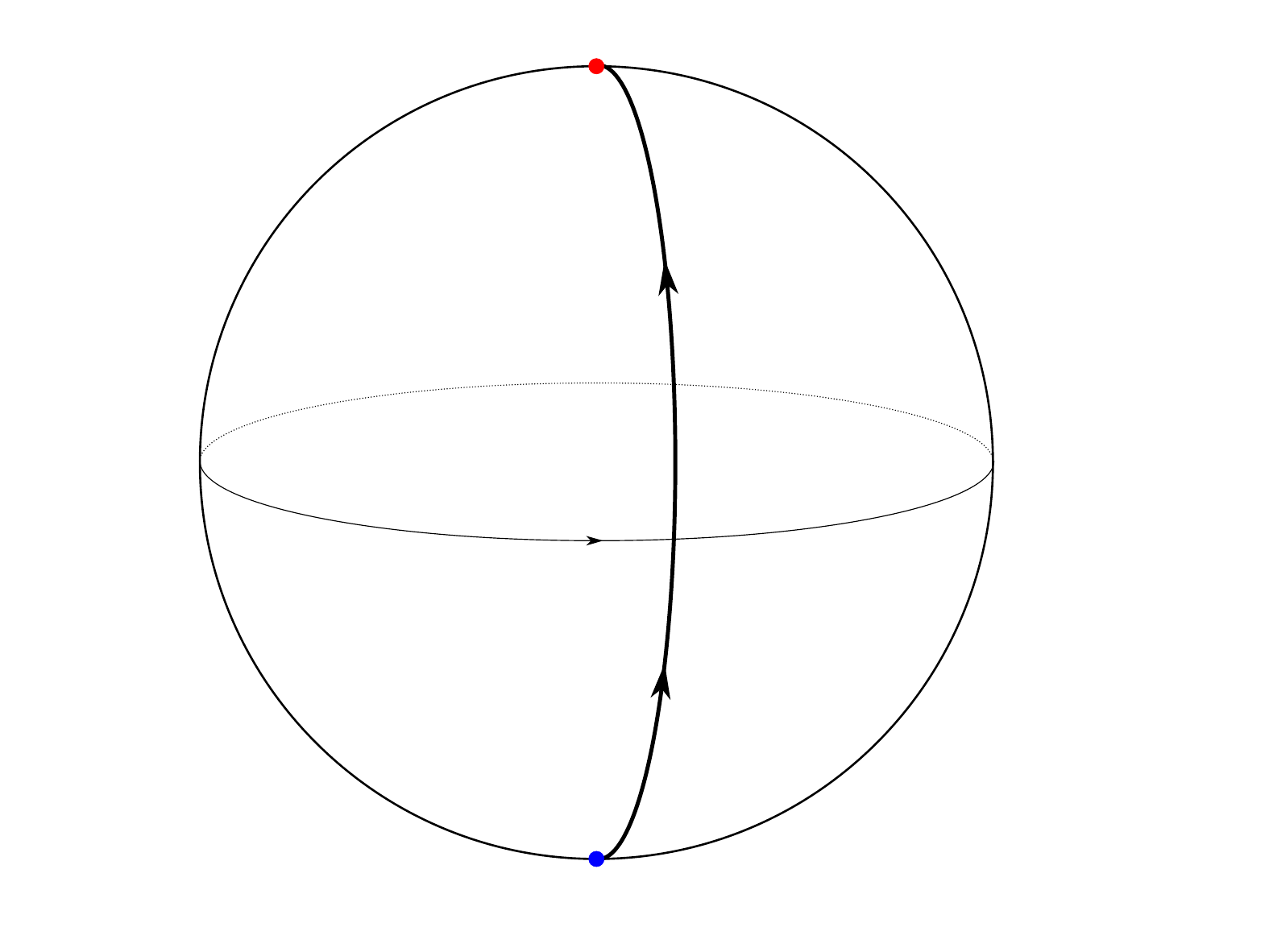}}%
    \put(0.43808345,0.71993294){\makebox(0,0)[lt]{\lineheight{1.25}\smash{\begin{tabular}[t]{l}$\vec{\mu}_1$\end{tabular}}}}%
    \put(0.43642997,0.02709436){\makebox(0,0)[lt]{\lineheight{1.25}\smash{\begin{tabular}[t]{l}$\vec{\mu}_2$\end{tabular}}}}%
    \put(0.80240115,0.37734537){\makebox(0,0)[lt]{\lineheight{1.25}\smash{\begin{tabular}[t]{l}$\hat{R}(1)$\end{tabular}}}}%
    \put(0,0){\includegraphics[width=\unitlength,page=2]{caseAroundX.pdf}}%
    \put(0.03712524,0.37792292){\makebox(0,0)[lt]{\lineheight{1.25}\smash{\begin{tabular}[t]{l}$\hat{R}(3)$\end{tabular}}}}%
    \put(0,0){\includegraphics[width=\unitlength,page=3]{caseAroundX.pdf}}%
  \end{picture}%
\endgroup%

    &
	\def\svgwidth{0.45\columnwidth}
\begingroup%
  \makeatletter%
  \providecommand\color[2][]{%
    \errmessage{(Inkscape) Color is used for the text in Inkscape, but the package 'color.sty' is not loaded}%
    \renewcommand\color[2][]{}%
  }%
  \providecommand\transparent[1]{%
    \errmessage{(Inkscape) Transparency is used (non-zero) for the text in Inkscape, but the package 'transparent.sty' is not loaded}%
    \renewcommand\transparent[1]{}%
  }%
  \providecommand\rotatebox[2]{#2}%
  \newcommand*\fsize{\dimexpr\f@size pt\relax}%
  \newcommand*\lineheight[1]{\fontsize{\fsize}{#1\fsize}\selectfont}%
  \ifx\svgwidth\undefined%
    \setlength{\unitlength}{453.54330709bp}%
    \ifx\svgscale\undefined%
      \relax%
    \else%
      \setlength{\unitlength}{\unitlength * \real{\svgscale}}%
    \fi%
  \else%
    \setlength{\unitlength}{\svgwidth}%
  \fi%
  \global\let\svgwidth\undefined%
  \global\let\svgscale\undefined%
  \makeatother%
  \begin{picture}(1,0.75)%
    \lineheight{1}%
    \setlength\tabcolsep{0pt}%
    \put(0,0){\includegraphics[width=\unitlength,page=1]{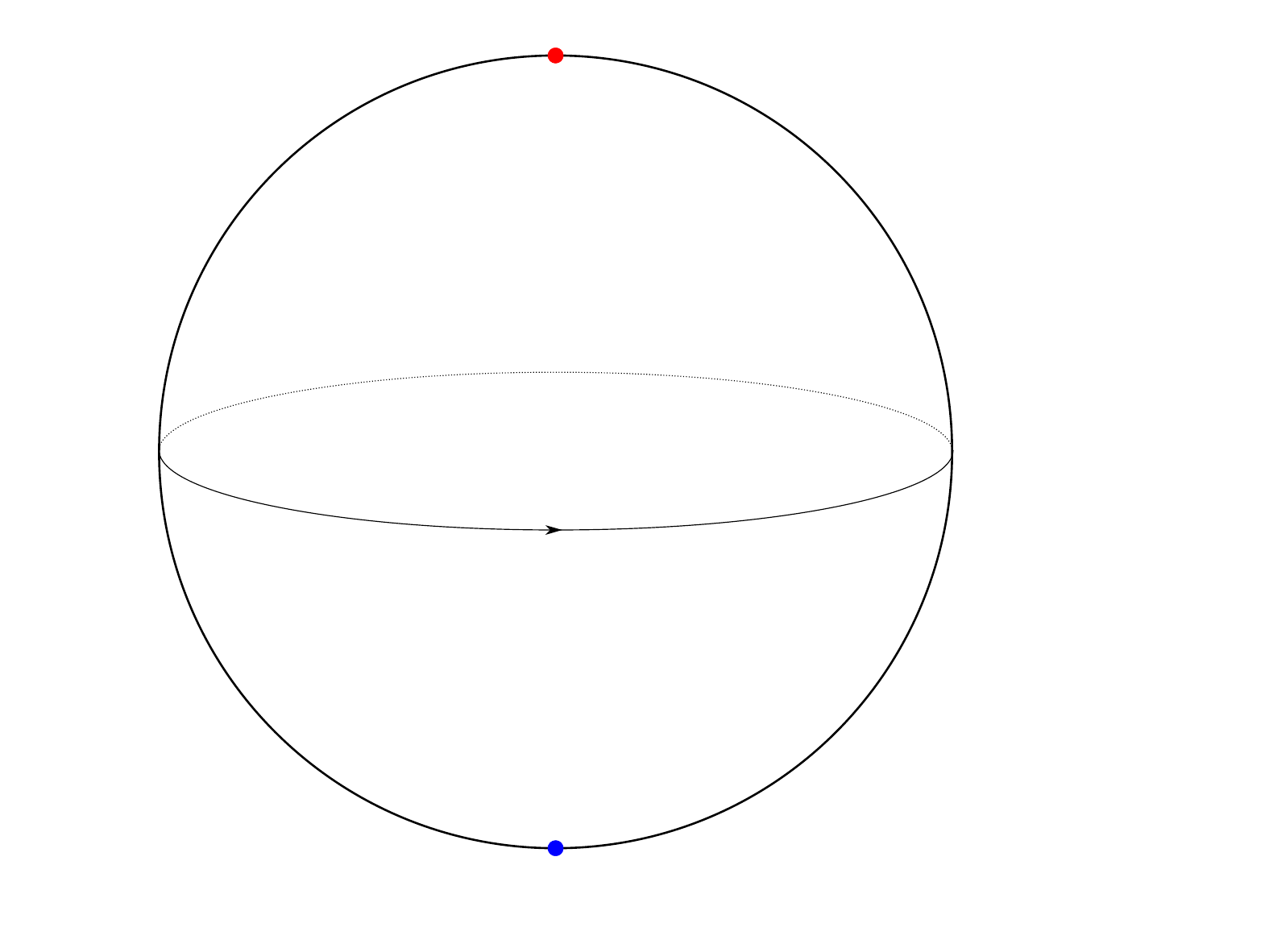}}%
    \put(0.40023814,0.72654743){\makebox(0,0)[lt]{\lineheight{1.25}\smash{\begin{tabular}[t]{l}$\vec{\mu}_1$\end{tabular}}}}%
    \put(0.40666438,0.03749418){\makebox(0,0)[lt]{\lineheight{1.25}\smash{\begin{tabular}[t]{l}$\vec{\mu}_2$\end{tabular}}}}%
    \put(0.76591769,0.38658991){\makebox(0,0)[lt]{\lineheight{1.25}\smash{\begin{tabular}[t]{l}$\hat{R}(1)$\end{tabular}}}}%
    \put(0,0){\includegraphics[width=\unitlength,page=2]{caseAroundY.pdf}}%
    \put(0.02389606,0.3838602){\makebox(0,0)[lt]{\lineheight{1.25}\smash{\begin{tabular}[t]{l}$\hat{R}(3)$\end{tabular}}}}%
    \put(0,0){\includegraphics[width=\unitlength,page=3]{caseAroundY.pdf}}%
  \end{picture}%
\endgroup%

    \end{tabular}
    \caption{The left figure illustrates the sequence of moves of Figs.~\ref{Bifurcation-X-notunwind} and \ref{BasicX-unwind}, with consecutive moves of $\vec{\mu}_2$ around the vertical meridian and movements of $\hat{R}$ around the equator.  
    The right figure illustrates the moves of Figs.~\ref{Bifurcation-Y-notunwind} and \ref{Bifurcation-Y-unwind}, with Figs.~\ref{Bifurcation-Y-notunwind} involving $\vec{\mu}_2$ circling the green meridian both times and the Fig.~\ref{Bifurcation-Y-unwind} involving $\vec{\mu}_2$ circling the green and then the orange meridian.}
    \label{fig:case34}
\end{figure}

In the remaining cases, Figs.~\ref{Bifurcation-X-notunwind}-\ref{Bifurcation-Y-notunwind}, 
$\vec{\omega}_1=0$ and the red dot stays put at the North pole. 
Clearly, for actual continuous spin and orbital motions there is no difference between $\vec{\omega}_2$ directed along $x$- or along  $y$-axis, due to the rotational symmetry.  In the table-top demonstrations, however, the motions are consecutive and the result depends on their order.  For example, Fig.~\ref{Bifurcation-X-notunwind} movements in terms of the direction two-sphere on the left of Fig.~\ref{fig:case34}, amount to the blue point circling along the vertical meridian (winding=-1), then the original black dot  $\hat{R}(1)$ moving along the equator to $\hat{R}(3)$ (winding=-1+1), the blue point circles the meridian again (with $\hat{R}$) on the other side (winding=-1+1+1), and the blue point completes its journey around the equator (winding=-1+1+1+1).

The movements of Fig.~\ref{BasicX-unwind}, on the other hand, consist of the blue point circling the meridian twice (winding=-2) followed by the black dot circling the equator (winding=-2+2).

The case of $\vec{\omega}_2$ aligned with the $y$-axis presents a bifurcation, since in this case the blue point passes through the black point.  This is the moment of bifurcation. Movements of Fig.~\ref{Bifurcation-Y-notunwind} (with $\vec{\omega}_2=\omega_2 \hat{y}+\epsilon \hat{x}$) from the direction two-sphere point of view are not so different from those of Fig.~\ref{Bifurcation-X-notunwind} discussed above.  The only difference is that the vertical meridian is rotated to the green meridian in Fig.~\ref{fig:case34}. The winding calculation remains the same.

Movements of Fig.~\ref{Bifurcation-Y-unwind} result in blue point circling the green meridian (winding=-1), black point moving along the equator from $\hat{R}(1)$ to $\hat{R}(2)$ (winding=-1+1), then the blue point circling the orange meridian (winding=-1+1-1), and the black point completing its trip around the equator (winding=-1+1-1+1).

\subsection{Twisting rate of the magnetic tube}
If we keep the central line fixed and rotate the first disk in Fig.~\ref{fig:tube} around its axis once, the new green path will wind by $2\pi$ relative to the original green path.  Similarly, rotating the second disc once winds this path by $-2\pi$. All the action in these cases is taking place in the cylinder in $\mathbb{F}=T^1S^2$ that is the preimage of the blue direction path.

Another important effect is due to the orbital motion, namely the point $\hat{R}=\vec{R}/|\vec{R}|$ is rotating around the equator with frequency $\Omega.$  Let us first focus on the path of directions in the sphere  $S^2$ while holding its origin $\vec{\mu}_1$ and its end $\vec{\mu}_2$ fixed.  
Moreover, we presume that $\vec{\mu}_1$ is in the Northern hemisphere and $\vec{\mu}_2$ is in the Southern hemisphere.
As discussed in Sec.~\ref{sec:F}, the path of directions passes in vicinity of $\hat{R}$, as illustrated in Fig.~\ref{fig:dirs}.  
As a result, during one orbital rotation the path of directions sweeps the whole surface of the sphere, and the magnetic tube twisting angle will increase   by $4\pi$ with each such sweep. 

Since the main phenomenon we are describing is topological in nature, we can  simplify our path of directions by deforming it to be the major arc from $\vec{\mu}_1$ to $\hat{R}$ followed by the major arc from $\hat{R}$ to $\vec{\mu}_2$.  
(For example we can deform the blue path of directions to the arc path just described, without passing through the antipodal point $-\hat{R}.$) 
During this deformation the covering path (dashed blue) will deform as well.  
In fact, we deform it so that at $\hat{R}$ the tangent vector is pointing due North.  Any such choice does not change the twisting, since rotating the tangent vector at $\hat{R}$ by some angle $\phi$ twists one part of the tube by  $-\phi,$ while twisting the other by  $\phi,$ thus not changing the total path twisting at all. 

This deformation makes it very easy to track the tube winding and its time evolution.  In fact, now that the frame at  $\hat{R}$ is fixed, we have a separate notion of twisting of the first part of the path and of its second part.  In particular, if $\vec{\mu}_1\in U_N$ and $\vec{\mu}_2\in U_S$, then as $\hat{R}$ circumnavigates the equator, the Northern part winds by $2\pi$, and so does the Southern part. Let us put this more geometrically.  When the point  $\hat{R}$ rotates around $\vec{\mu}_1$ {\em counterclockwise}, the $\vec{\mu}_1-\hat{R}$ path segment is twisted by $2\pi.$ Similarly, whenever,  $\hat{R}$ moves around $\vec{\mu}_2$ {\em clockwise} the segment $\hat{R}-\vec{\mu}_2$ is twisted by $2\pi.$ (Note, the difference is due to the opposite orientation:  $\vec{\mu}-\hat{R}$ vs $\hat{R}-\vec{\mu}$.)

Now, as the path of directions returns to itself after one orbital rotation, what happens to the magnetic flux tube? It is twisted by $4\pi$ if  $\vec{\mu}_1$ is North of equator and $\vec{\mu}_2$ is South of it. It is twisted by $-4\pi$ if  $\vec{\mu}_1$ is South of equator and $\vec{\mu}_2$ is North of it. And it is not twisted at all if $\vec{\mu}_1$ and $\vec{\mu}_2$ are on the same side of the equator.   This effect of the orbital motion is due to the fact that the bundle $\mathbb{F}$ of circles over the sphere  has Hopf number 2.

Of course in general the disks will spin and orbit each other at the same time. 
Consider the first disk rotating with angular velocity $\vec{\omega}_1$ and the second with angular velocity $\vec{\omega}_2.$ Then, the points $\vec{\mu}_1$ and $\vec{\mu}_2$ will be in circular motions around $\hat{\omega}_1=\frac{\vec{\omega}_1}{\omega_1}$ and $\hat{\omega}_2=\frac{\vec{\omega}_2}{\omega_2}$ respectively. 
Let $\alpha_1$ be the angle between  $\vec{\Omega}$ and  $\vec{\omega}_1$ and let $\beta_1$ be the angle between  $\vec{\omega}_1$ and $\vec{\mu}_1$, as in Fig.~\ref{fig:trigonom}. 
Similarly, we define $\alpha_2$ and $\beta_2$ using $\vec{\omega}_2$ and $\vec{\mu}_2$. 
Now, as the first disk undergoes one such rotation around $\omega_1$ 
it rotates around its axis once, therefore, as $\vec{\mu}_1$ undergoes 
one rotation around $\frac{\omega_1}{|\omega_1|}$ the circle fiber of $\mathbb{F}$ above it rotates once (relative to the fiber at $\hat{\omega}_1$).
The resulting twisting, however, depends on whether $\vec{R}$ lies inside or outside the smaller disk bounded by the circle traversed by $\omega_1$ (or, respectively, $\omega_2$).

\subsection{Effective rate of twisting}
\begin{figure}[!ht]
    \centering
	\def\svgwidth{1\columnwidth}
 	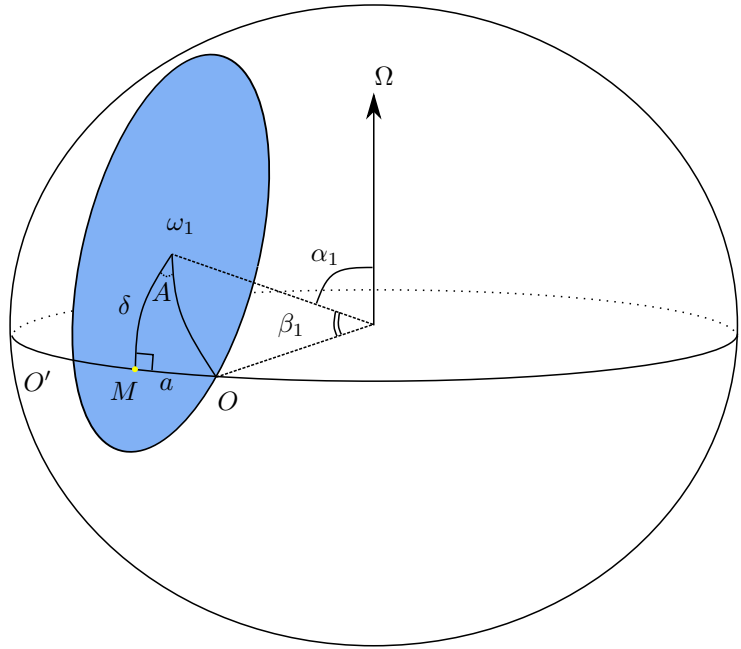
 
    \caption{The blue spherical disc is centered at $\hat{\omega}_1$ and bounded by the trajectory of $\hat{\mu}_1.$ Its opening angle is $\beta_1.$  The spherical triangle $\triangle \omega_1 M O$ supports half of the equatorial arch  between the two points $O$ and  $O'$ where the orbit of $\hat{\mu}_1$ crosses the equator. The angle $\angle O\omega_1M$ is denoted by $A$, while the length of its opposing side $MO$ is  $a.$}
    \label{fig:trigonom}
\end{figure}

If the motion of $\vec{\mu}_1$ is happening entirely in one hemisphere (either Northern or Southern) and the same holds for $\vec{\mu}_2,$ then the twisting rate 
of the $\hat{\mu}_1-\hat{R}$ part is
$- \omega_1\, \mathrm{sign}\,\vec{\omega}_1\cdot \vec{\mu}_1  
+ \Omega\, \mathrm{sign} \vec{\Omega}\cdot\vec{\mu}_1$,
while the twisting rate of the $\hat{R}-\hat{\mu}_2$ part is 
$- \omega_2\, \mathrm{sign}\,\vec{\omega}_2\cdot \vec{\mu}_2
	+ \Omega\, \mathrm{sign} \vec{\Omega}\cdot\vec{\mu}_2.$ 
Assembling the two together we have the rate of twisting of the whole path:
\begin{align}
	- \omega_1\, \mathrm{sign}\,\vec{\omega}_1\cdot \vec{\mu}_2
	+ \omega_2\, \mathrm{sign}\,\vec{\omega}_2\cdot \vec{\mu}_2
	+ \Omega\, (\mathrm{sign} \vec{\Omega}\cdot\vec{\mu}_1 -\mathrm{sign} \vec{\Omega}\cdot\vec{\mu}_2)
.\end{align}
The above holds so long as $\alpha_i+\beta_i< \frac{\pi}{2}$ or after $\alpha_i-\beta_i>\frac{\pi}{2}$ for $i=1,2$. Otherwise the circle on the direction sphere that $\vec{\mu}_i$ traces out dips in the other hemisphere.  What fraction of this circle is in the other hemisphere? 
In other words, how long is the arc between the two points $O$ and  $O'$ (see Fig.~\ref{fig:trigonom}) at which $\vec{\mu}_1$ passes the equator?

Let $\delta_1=\frac{\pi}{2}-\alpha_1$ be the latitude of $\vec{\omega}_1.$ Then
from the spherical triangle formed by $\vec{\omega}_1,$ $O,$ and the equator midpoint $M$ between  $O$ and  $O'$ we have 
 \begin{align}
	 \cos \beta_1 & = \cos a \cos \delta_1,\\
	 \sin A & = \frac{\sin a}{\sin \beta_1}
.\end{align}
Thus $\sin A = \frac{\sqrt{\sin^2 \alpha_1 - \cos^2 \beta_1}}{\sin \alpha_1 \sin \beta_1},$ and the fraction of the $\vec{\mu}_1$ circle that is in the opposite hemisphere is 
\begin{align}\label{Eq:dip}
	f(\alpha,\beta) & =
	\begin{cases}
	\frac{2 A}{2\pi}= \frac{1}{\pi} \arcsin \frac{\sqrt{\sin^2 \alpha-\cos^2 \beta}}{\sin \alpha \sin \beta}& \text{for }  
	\frac{\pi}{2}-\beta<\alpha<\frac{\pi}{2}+\beta\\
        0 & \text{otherwise}
	\end{cases}
.\end{align}

Assuming for a moment that the rotation frequencies are much higher than the orbital frequency: $\omega_1, \omega_2\gg \Omega,$ the average twisting rate $\mathcal{R}$  is 
\begin{align}
\mathcal{R}=	- (1-2f_1)\omega_1\, \mathrm{sign}  ( \vec{\omega}_1\cdot\vec{\mu}_1 ) 
	+\Omega\, \mathrm{sign}(\,\vec{\Omega}\cdot \vec{\omega}_1) \, \mathrm{sign}\,({\vec{\mu}_1}\cdot \vec{\omega}_1)
	\nonumber\\
	+ (1-2f_2)\omega_2\, \mathrm{sign} (\vec{\omega}_2\cdot\vec{\mu}_2 )
	-\Omega\, \mathrm{sign}\, ( \vec{\Omega}\cdot \vec{\omega}_2) \, \mathrm{sign}\,({\vec{\mu}_2}\cdot \vec{\omega}_2)
.\end{align}
Here $f_j=f(\alpha_j,\beta_j)$ given by Eq.~\ref{Eq:dip} with $\alpha_j$ being the angle between $\vec{\Omega}$ and $\vec{\omega}_j$ and $\beta_j$ angle between  $\vec{\omega}_j$ and  $\vec{\mu}_j.$

\begin{figure}[!ht]
\centering
\includegraphics[width=.99\textwidth]{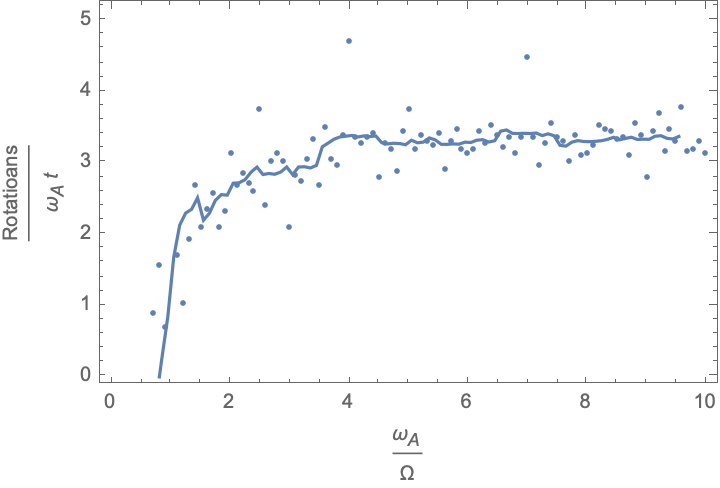}
\caption{ Rotations per unit time as function of $\omega_A/\Omega$,  $\theta_A =\pi/3$  and $\theta_{A,c} = \pi/4$; solid line is the running average. At large $\omega_A/\Omega \geq 1$ a finite limit is reached.  Both the average curve and the variations carry information about parameters of the system.  }
\label{Rotationspertime} 
\end{figure}

This provides the average rate of winding of the magnetosphere for the general relative orientation of spins and orbital angular velocity.  The averaging used applies whenever $\omega_1,\omega_2 \gg \Omega.$ Fig.~\ref{Rotationspertime} illustrating the time average converging to $\mathcal{R}.$ 
Whenever the spins and angles are such that this rate $\mathcal{R}$ is low, a slow winding of the magnetosphere is possible on average over time.  This does not imply, however, that these configurations allow for gradual storage of energy in the magnetosphere, since the critical winding might/is likely to be breached in the interim. 
With this in mind we focus on identifying other possible resonances.

\begin{figure}[h!]
\centering
\includegraphics[width=.99\textwidth]{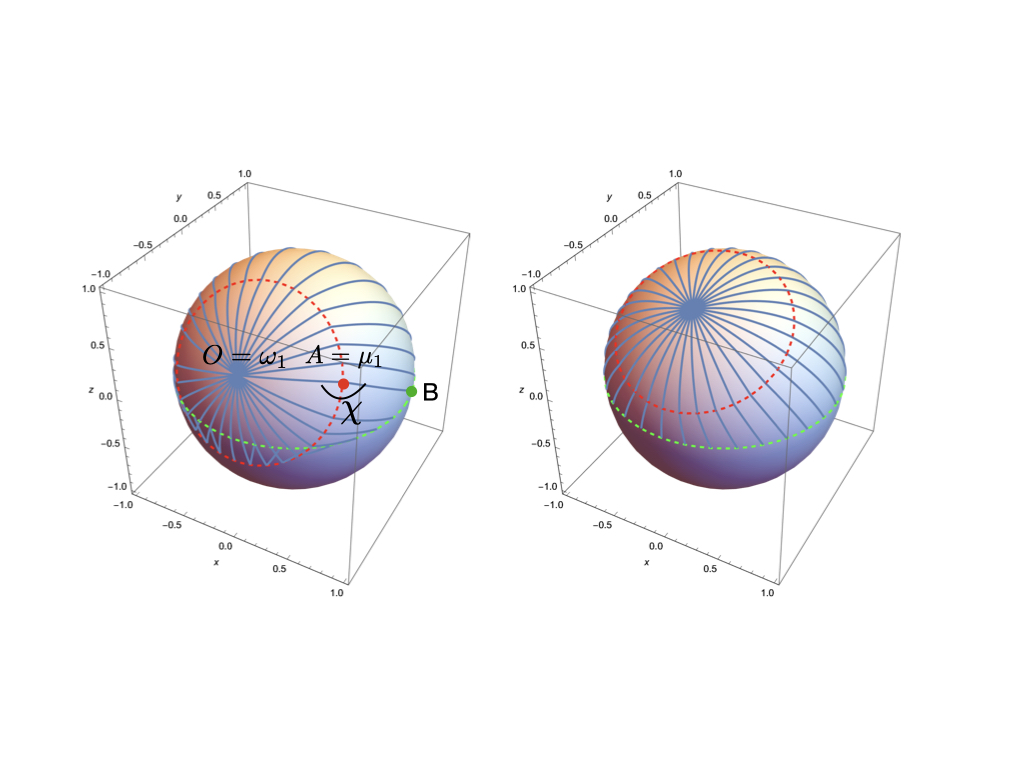}
\caption{Example of twisting if/when the point $A$ crosses the equator. The sphere represents a space of the directions of the magnetic field. Red dashed circle is the trajectory of the direction of the  magnetic field at point $A$ (direction of magnetic moment of $1$); it rotates with the angular frequency $\omega_A=\omega_1$ around point $O$, the direction of the spin of the first star.
For both plots $\omega_1= \Omega$: one spin per one full orbital rotation. Left panel inclination $\theta_A =\pi/3$, right panel  $\theta_A =\pi/6$ (these are the angles between orbital momentum $\vec{\Omega}$ and the spin $\vec{\omega}_1$. Radius of the circle $\theta_{A,c} = \pi/4$ (this is the angle between the spin $\vec{\omega}_1$ and the magnetic field $\vec{\mu}_1$). Notice how arcs cross-over in the lower part in the left panel, when point $A$ is below the equator. To trace the twist, we follow the value $\chi$ of the angle $\angle{OAB}$ in Fig. \ref{Rotations}.
 }
\label{Divingunder} 
\end{figure}

\begin{figure}[!ht]
\centering
\includegraphics[width=.99\textwidth]{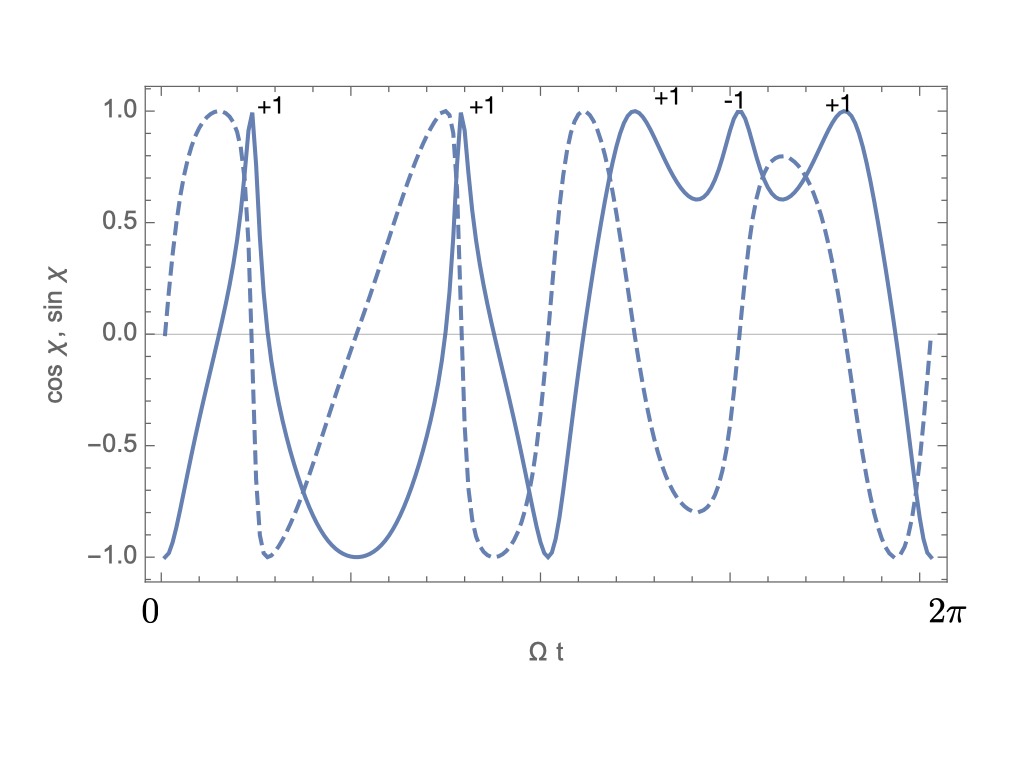}
\caption{Counting the twist. Plotted is the angle $\chi$, spin of A - magnetic moment of A - direction to B, ($\cos \chi$ - solid line,  $\sin \chi$ - dashed line). One rotation is added when $\cos \chi=1 $ and  $\sin \chi$ changes from positive to negative; one rotation is subtracted when $\sin \chi$ changes from negative to positive. In this example $\omega_A = 5\Omega$,   $\theta_A =\pi/3$  and $\theta_{A,c} = \pi/4$.}
\label{Rotations} 
\end{figure}

\subsection{Importance of the phase}

We can have multiple  non-winding resonances besides the 2:1 resonance; but their appearance depends on the phase.  Thus, in a given system one can observe several various resonances. These higher resonances however are much less stable.

Consider a snapshot with $\omega_1, \hat{R}=M,$ and $\vec{\mu}_1$ aligned {\bf lying on} a spherical radius of the blue disk in Fig.~\ref{fig:trigonom}. 
If in the near future $\hat{R}$ reaches point $O$ before  $\vec{\mu}_1$ does, then it also implies that in the recent past $\vec{\mu}_1$ passed through point $O'$ before  $\hat{R}$ did.  In other words, the motion of $\vec{\mu}_1$ over the arch $O'O$ canceled the effect of  $\hat{R}$ traversing the equator. (I.e. $\vec{\mu}_1$ made one clockwise rotation around $\hat{R}$.) 

The necessary condition for such occurrence is 
$\frac{a}{\Omega}<\frac{A}{\omega}$, 
which translates to 
\begin{align}
\omega \arccos \frac{\cos\beta}{\sin\alpha}<\Omega \arcsin \frac{\sqrt{\sin^2\alpha-\cos^2\beta}}{\sin\alpha \sin\beta} 
.\end{align}

If the above snapshot occurs and the ration of the two frequencies is rational, the above snapshot will recur with regularity. E.g. if $\frac{\omega_1}{\Omega}=\frac{2}{3},$ then every two orbital periods $\vec{\omega}_1, \hat{R},$ and $\vec{\mu}_1$ are aligned again and the  orbital winding effect is compensated by the above maneuver. Therefore, the effective magnetic tube winding rate is $\omega_1-\Omega(1-\frac{1}{3})$, which vanishes. In particular, if it so happens that $\vec{\mu}_2$ is engaged in a similar resonant configuration, then the total winding rate is zero.  To summarize, we have another resonant configuration
\begin{align}
	\omega_1&=\omega_2=\frac{2}{3}\Omega,&
	\omega_1+\omega_2=\frac{4}{3}\Omega
.\end{align}
In contrast to $2:1$ resonance which was very stable with respect to changes of angles, phases, and frequencies, this $4:3$ resonance is sensitive not only to angles and frequencies, but even to the phase, i.e. to the particular momentary (approximate) alignment.  Nevertheless, during the binary evolution such resonances can occur. {\bf Moreover, there are many other possible resonances.} If they are observed, they can provide detailed information about the individual spins, their orientations, and about the orbital frequency.

\begin{figure}[!ht]
    \centering
	\def\svgwidth{0.7\columnwidth}
\begingroup%
  \makeatletter%
  \providecommand\color[2][]{%
    \errmessage{(Inkscape) Color is used for the text in Inkscape, but the package 'color.sty' is not loaded}%
    \renewcommand\color[2][]{}%
  }%
  \providecommand\transparent[1]{%
    \errmessage{(Inkscape) Transparency is used (non-zero) for the text in Inkscape, but the package 'transparent.sty' is not loaded}%
    \renewcommand\transparent[1]{}%
  }%
  \providecommand\rotatebox[2]{#2}%
  \newcommand*\fsize{\dimexpr\f@size pt\relax}%
  \newcommand*\lineheight[1]{\fontsize{\fsize}{#1\fsize}\selectfont}%
  \ifx\svgwidth\undefined%
    \setlength{\unitlength}{576bp}%
    \ifx\svgscale\undefined%
      \relax%
    \else%
      \setlength{\unitlength}{\unitlength * \real{\svgscale}}%
    \fi%
  \else%
    \setlength{\unitlength}{\svgwidth}%
  \fi%
  \global\let\svgwidth\undefined%
  \global\let\svgscale\undefined%
  \makeatother%
  \begin{picture}(1,1.13020833)%
    \lineheight{1}%
    \setlength\tabcolsep{0pt}%
    \put(0,0){\includegraphics[width=\unitlength,page=1]{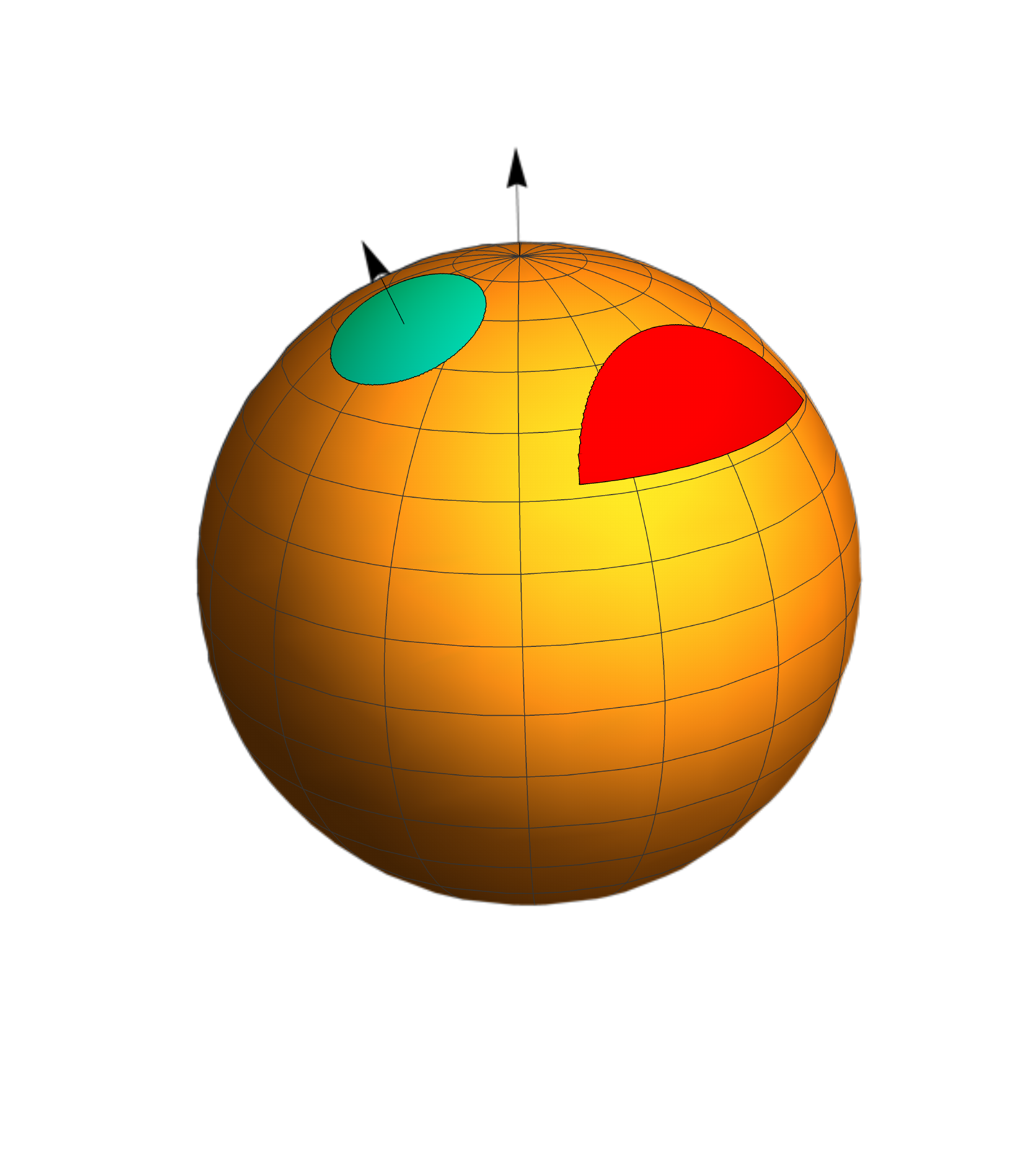}}%
    \put(0.51697226,0.99650861){\makebox(0,0)[lt]{\lineheight{1.25}\smash{\begin{tabular}[t]{l}$\omega$\end{tabular}}}}%
    \put(0.32622734,0.92520209){\makebox(0,0)[lt]{\lineheight{1.25}\smash{\begin{tabular}[t]{l}$\mu$\end{tabular}}}}%
  \end{picture}%
\endgroup%

    \caption{ The magnetic cap, centered around $\hat{\mu}$ in cyan and the resonant region in red for $\lambda=\frac{\pi}{3}$. A priori, the magnetic cap can be positioned anywhere on the sphere of directions.  It is the overlap of the magnetic sphere and the resonant region that determines the part of the magnetosphere involved in the 4:3 resonance.}
    \label{fig:43res}
\end{figure}

Let us estimate the local region of magnetosphere that can contribute to the 4:3 resonance.  The $\hat{R}$ transit time (between $O'$ and  $O$) is  $T_R=\frac{2a}{\Omega}$.  The angular distance traveled by $\hat{B}$ during this time is $\omega T_R=\frac{\omega}{\Omega} 2 a.$
The remaining angular span of the $O'O$ arc (on the boundary of the blue spherical disk in Fig.~\ref{fig:trigonom}) is therefore $2A-\frac{\omega}{\Omega}2a$. This is the arc of the $\hat{B}$ points involved in the above described behavior leading to the 4:3 resonance.  I.e. a point $\hat{B}$ passes $O'$ before  $\hat{R}$ does, and then $\hat{B}$ passes 
$O$ after  $\hat{R}$ does. Thus, for each value of $\beta$ there is an arc of angular size 
 \begin{align}
        2A-\frac{2\omega}{\Omega}a = 2 \arcsin \frac{\sqrt{\sin^2\alpha - \cos^2\beta}}{\sin\alpha \sin\beta} - 2 \frac{\omega}{\Omega} \arccos \frac{\cos\beta}{\sin \alpha} 
.\end{align}
The relevant solid angle formed by the magnetic line star surface directions involved in the resonance is
\begin{align}
        \int_{\frac{\pi}{2}-\alpha}^{\arccos \frac{\sin\alpha}{2}} (\sin\beta) (2A-2\frac{\omega}{\Omega}a) d\beta
.\end{align}
The upper limit is reached when the equatorial arc $O'O$ spans one third of the equator, i.e.  $2a<\frac{2\pi}{3}$, implying $\cos\beta<\frac{1}{2}\sin\alpha$.

Now the fragility of this resonance can be illustrated in Fig.~\ref{fig:43res}.  The resonant region, indicated in red, involves only those magnetic lines whose direction at the star surface lie in both the interacting magnetic polar cap (shown in cyan) and in the resonant (red) region. 
With our formulas above one can estimate how likely it is that these two regions intersect at all and what the maximal overlap of these two regions can be.

\section{What astrophysical observational  effects we expect: precursor flares in  merging neutron stars}
\label{merging}

Qualitatively, if only one star is magnetized, the corresponding slowly evolving  powers are \citep{2001MNRAS.322..695H,2011PhRvD..83l4035L}
\begin{equation}
L_{1M} \sim  \frac{G B_{{NS}}^2 M_{{NS}} R_{{NS}}^8}{c r^7}=3 \times { 10^{41} }{(-t_m)^{-7/4}}\, {\rm \, erg \, s^{-1}}
\label{L1}
.\end{equation}
(Index $1M$ indicates here that the interaction is between single magnetized neutron star and unmagnetized one.)
The time to merger  $t_m$ 
\begin{equation}
- t_m= \frac{5}{256} \frac{c^5}{G^3} \frac{r^4}{M_1 M_2 (M_1+M_2)}
\label{tm}
 \end{equation}
 is measured in seconds in Eq.  (\ref{L1}).

Magnetospheric interaction of two {\it magnetized} neutron stars can generate larger luminosity: the interactions is between two magnetospheres, with effectively larger radius
\citep{2019MNRAS.483.2766L}
\begin{equation}
L_{2M} \sim \frac{ B_{{NS}}^2 G  M_{{NS}} R_{{NS}}^6}{c  r^5}    =
\frac{c^{21/4} B_{{NS}}^2 R_{{NS}}^6}{
   (-t)^{5/4} ( G M_{NS})^{11/4}}=
 6 \times { 10^{42} }{(-t_m)^{-5/4}}\, {\rm \, erg \, s^{-1}}
    \label{L2}
.\end{equation}
(Index $2M$ indicates here that the interaction is between two magnetized neutron star.)
Thus $L_2 \geq  L_1$ dominates  prior to merger.

Both powers (\ref{L1}-\ref{L2}) are not very large: they will be missed by high energy observatories, but may be sufficiently bright if a large fraction is emitted in radio, producing a signal  \citep{2019MNRAS.483.2766L}
 \begin{equation}
F_R \sim \eta_R  \frac{L_2}{4\pi d^2  \nu}\approx 0.1 {\rm \, Jy}\,  \eta_{R,-5} (-t)^{-5/4}
\label{FR}
,\end{equation}
where $\eta_R$ is a fraction of the energy going into radio.

Topological resonance, especially the global one (\ref{2to1}) allows two start to establish nearly permanent magnetic link.  A slow change of the parameters would allow storage of magnetic energy within these configurations, and the subsequent release of the stored energy when the magnetic fields\ become too sheared/too twisted. 

Equating neutron star beat-plus spins  with the orbital frequency  at separation $r$  before the merger
 \begin{equation}
\omega= \omega_1+\omega_2  =2  \sqrt{\frac{G (M_1+M_2)}{r^3}}
 ,\end{equation}
 and using time to merger (\ref{tm}),  the 2:1 resonance  occurs at 
 time
 \begin{equation}
t_{1-2}=(-t_m)=  10^{-3} \frac{ c^5 (M_1+M_2)^{1/3}}{G^{5/3} M_1 M_2 \omega^{8/3}}= 5\times 10^{5} P_s^{8/3} \, sec
,\end{equation}
where $P_s = 2\pi/\omega$ is the sum-beat (addition)   period of the NS spin's, $M_1$ and $M_2$ are masses of neutron stars (assumed to be equal to $1.4M_\odot$. 
Thus, the most interesting case if one of the neutron star is a recycled millisecond pulsar. For example, if $P_s = 10 $ msec, then $(-t_m)=2$ sec. 
At that moment stars are separated by 
\begin{equation}
r= ( G (M_1+M2))^{1/3} \Omega^{-2/3} \approx 10^7 {\rm cm},
\end{equation}
approximately ten times the radius.

We can also estimate how flaring will evolve with time. At exact resonance there is no shear, no flaring. As the orbit shrinks,  system  gets out of the resonance, field lines are becoming twisted. Assuming that flares occur after a fixed twist angle $\sim 1$, time between flares evaluates to
\begin{equation}
t_f= \frac{4 P}{3 \pi} \frac{t_{1-2}}{t-t_{1-2}}
\end{equation}

Next, we need to estimate the amount of connected magnetic field lines.
 Vacuum dipoles provide a good first approximation to the magnetospheric structure. A simple case is that of anti-aligned magnetic moment, both orthogonal to the orbital plane.  Let the two stars be located at $\{ x_{1,2}= \pm r/2 , y_{1,2} =0, z_{1,2}=0\}$.  In the plane $y=0$ (the plane that contains the vector connecting the stars and their magnetic moments) the total magnetic flux function is 
\begin{equation}
\Psi_{tot} = \frac{ (x-r/2)^2}{ \left( (x-r/2)^2 +z^2\right)^{3/2}} +  \frac{ (x+r/2)^2}{ \left( (x+r/2)^2 +z^2\right)^{3/2}} 
\end{equation}
($\Psi$ is constant on each field line. The inner separatrix between regions (i) and (ii) are given by $\Psi_{tot}  = 4/r$, while outer by $\Psi_{tot}  = 0$. 
At large distances the outer separatrix are at $45$ degrees.

  For two neutron stars of radius $R_{NS}$ separated by distance $r$ the angular size of the patch of connected field lines can be estimated as
  \begin{equation}
  \theta_{pc} \approx 2\sqrt{\frac{R_{NS}}{r}}
  .\end{equation}
  Thus, the energy in the connected magnetic field can be estimate as 
  \begin{equation}
E_{B,c} \approx \frac{B_{NS}^2 R_{NS}^4}{r}
.\end{equation}
Therefore, about one tenth of the total magnetic energy can be released in a flare.
For example, for merging neutron stars,
  \begin{equation}
E_{B,c} \approx0.08  \frac{B_{NS}^2 R_{NS}^4 \omega^{2/3}}{(G M_{NS})^{1/3}} =
10^{41} B_{12}^2 P_{s,-2}^{-2/3} \, {\rm erg}
,\end{equation}
for a  neutron star with a period  $ P=10^{-2} P_{s,-2}$ seconds.
If reconnection occurs on light travel time over orbital separation, the expected power is
\begin{equation}
L _{B,c} =\frac{ E_{B,c}}{r/c} = 10^{-1}  \frac{B_{NS}^2  c R_{NS}^4 \omega^{4/3}}{(G M_{NS})^{2/3}} =
3 \times 10^{44}  B_{12}^2 P_{s,-2}^{-4/3}  \, {\rm erg\, s}^{-1}
.\end{equation}
This is  a mild  amount of energy/mild luminosity even for fast-spinning msec neutron star. It  still can be detected from $\sim 100$ Mpc distances with all-sky high energy monitors, and also can be seen in targeted observations  (\eg\ due to preliminary LISA localization).  Attempts to detect the precursor emission have been discussed by \cite{2019ApJ...877L..39C,2020ApJ...905L..25S}.

\section{Discussion}

In this work we consider topological structure of magnetically interacting binaries. We are particularly interested in configurations whose magnetospheres wind slowly - when the effects of spins and orbital motion (periodically) compensate each other. 
We point out that beside the  very restricted cases of fully locked rotation and equal antiparallel spin in the orbital frame, 
there are other slow-winding configurations that can unwind locally  (in a sense that a special set of magnetic tubes may wind slowly, while other magnetic tubes keep winding at high rate). 
The most interesting case is when beat-plus frequency $\omega_1 +\omega_2$ equals two times the orbital frequency.  This globally non-winding configuration is achieved in  a broad range of parameters (relative directions of spins, magnetic moments and orbital angular momentum) that we identified.  

There are no other globally unwinding configurations beside the tree cases mentioned above. But there can be a specific magnetic tubes (with slow winding)  connecting the two stars that are slowly winding, while other field are are getting twisted. 
Whenever the fraction of such tubes is significant, one might expect gradual energy transfer from rotational to magnetic energy and its eventual release.

Our main mathematical  perspective is viewing a magnetic tube as a line in the space $\mathbb{F}$ of unit tangent circles to the two-sphere of directions.  This makes apparent that the bifurcation point corresponds to one of the magnetic moments being (nearly) aligned with the binary separation vector: $\vec{\mu}_1$ or  $\vec{\mu}_2 \parallel \vec{R}$.
The tube winding is due to 1) the spins twisting its two ends and 2) the orbital motion contributing one full twist to each half as $\hat{R}$ orbits around $\hat{\mu}_j$ in the sphere of directions.
The latter effect originates in the topology of $\mathbb{F}$, which is a degree 2 Hopf fibration.  It is this fact that ensures persistence and relative stability of the 2:1 resonance.

The main predicted phenomena for the case of merging neutron stars are the precursor flares, which can occur just few seconds before the main gravitational wave event \citep{2001MNRAS.322..695H,2019MNRAS.483.2766L}. The flare luminosity can be higher than the slowly varying persistent one; it is  also  easier to detect flaring events.

Curiously,  since magnetic unwinding for other (not 2:1) resonances depends on phase, it may not occur every orbit, but with some periodicity. Analyzing the periodicity may constrain the absolute values of the spins.  Generally, unwinding  
 2:1 resonance is very stable. Yet if higher order resonances are observed, a lot of detailed information may be inferred.   Higher order resonances may have similar power to the basic one, Fig. \ref{fig:43res}.

A few further  modifications  are planned. Elliptical orbits will add another complication: non-constant rotation of vector ${\bf R}$/point $B$ which  would rotate with  changing Keplerian angular velocity.
Our results, however, are essentially topological and thus should not be  sensitive to such modifications. 

Our perspective also calls for refinements, one can use it to obtain the rate of magnetic tube slow winding (without the averaging) and the rate at which the resulting energy is released (whenever the magnetic tube winding exceeds some critical value).  One can also estimate the dependence of the resulting energy release on the spin and orbital rotation parameters as well as its time signatures.  All this is left for future exploration.
 
In  application to  exoplanet-star magnetic interactions, few comments are due. First, magnetic coupling between magnetospheres can occur only within the Alfven  radius of the parent star ($\sim 10 R_\odot$ for the Sun).  Tidal effects are likely to bring the system into corotation \citep{1981A&A....99..126H} 
 \citep[though more complicated dynamics like tidal spin-ups is also been considered][]{2021arXiv210705759T}.  For $z$-aligned spins this would make the magnetospheric structure in the rotating frame been static.  
 On the other hand, 
there are evidence of cases of high obliquity, when the orbit of the star is highly inclined with respect to stellar spin \citep[see][for review]{2015ARA&A..53..409W}. 
What is still required is alignment of at least one spin with the orbital angular momentum.
 Finally, there are indeed observations of the magnetospheric interactions  of planet-modulated chromospheric and radio emission \citep{2019NatAs...3.1128C,2020NatAs...4..577V,2021A&A...645A..59T}.

  \section{Acknowledgements}

The work of ML is  supported by 
NASA grants 80NSSC17K0757 and 80NSSC20K0910,   NSF grants 1903332 and  1908590.
The work of SCh is supported by the Charles Simonyi Endowment at the Institute for Advanced Study. 

We would like to thank Brad Hansen and  Matt Shultz for discussions.

\section{Data availability}
The data underlying this article will be shared on reasonable request to the corresponding author.

 \bibliographystyle{apj} 
  \bibliography{BibTex} 

\begin{thebibliography}{25}
\expandafter\ifx\csname natexlab\endcsname\relax\def\natexlab#1{#1}\fi

\bibitem[{{Antoine}(2021)}]{2021arXiv210405968A}
{Antoine}, S. 2021, arXiv e-prints, arXiv:2104.05968

\bibitem[{{Arnold}(1974)}]{Arnold74}
{Arnold}, V.~I. 1974, Proc. Summer School in Differential Equations, Erevan.
  Armenian SSR A d Sci.

\bibitem[{Arnold(1986)}]{Arnold73}
Arnold, V.~I. 1986, Selecta Math. Soviet., 5, 327, selected translations

\bibitem[{{Bildsten} \& {Cutler}(1992)}]{1992ApJ...400..175B}
{Bildsten}, L., \& {Cutler}, C. 1992, \apj, 400, 175

\bibitem[{{Buckley} {et~al.}(2017){Buckley}, {Meintjes}, {Potter}, {Marsh}, \&
  {G{\"a}nsicke}}]{2017NatAs...1E..29B}
{Buckley}, D.~A.~H., {Meintjes}, P.~J., {Potter}, S.~B., {Marsh}, T.~R., \&
  {G{\"a}nsicke}, B.~T. 2017, Nature Astronomy, 1, 0029

\bibitem[{{Callister} {et~al.}(2019){Callister}, {Anderson}, {Hallinan},
  {D'addario}, {Dowell}, {Kassim}, {Lazio}, {Price}, \&
  {Schinzel}}]{2019ApJ...877L..39C}
{Callister}, T.~A., {et~al.} 2019, \apjl, 877, L39

\bibitem[{{Cauley} {et~al.}(2019){Cauley}, {Shkolnik}, {Llama}, \&
  {Lanza}}]{2019NatAs...3.1128C}
{Cauley}, P.~W., {Shkolnik}, E.~L., {Llama}, J., \& {Lanza}, A.~F. 2019, Nature
  Astronomy, 3, 1128

\bibitem[{{Goldreich} \& {Julian}(1969)}]{GJ}
{Goldreich}, P., \& {Julian}, W.~H. 1969, \apj, 157, 869

\bibitem[{{Hansen} \& {Lyutikov}(2001)}]{2001MNRAS.322..695H}
{Hansen}, B.~M.~S., \& {Lyutikov}, M. 2001, \mnras, 322, 695

\bibitem[{{Hut}(1981)}]{1981A&A....99..126H}
{Hut}, P. 1981, \aap, 99, 126

\bibitem[{{Lyutikov}(2011)}]{2011PhRvD..83l4035L}
{Lyutikov}, M. 2011, \prd, 83, 124035

\bibitem[{{Lyutikov}(2019)}]{2019MNRAS.483.2766L}
---. 2019, \mnras, 483, 2766

\bibitem[{Moffatt \& Dormy(2019)}]{moffatt_dormy_2019}
Moffatt, K., \& Dormy, E. 2019, Self-Exciting Fluid Dynamos, Cambridge Texts in
  Applied Mathematics (Cambridge University Press)

\bibitem[{{Most} \& {Philippov}(2020)}]{2020ApJ...893L...6M}
{Most}, E.~R., \& {Philippov}, A.~A. 2020, \apjl, 893, L6

\bibitem[{{Palenzuela} {et~al.}(2013){Palenzuela}, {Lehner}, {Ponce},
  {Liebling}, {Anderson}, {Neilsen}, \& {Motl}}]{2013PhRvL.111f1105P}
{Palenzuela}, C., {Lehner}, L., {Ponce}, M., {Liebling}, S.~L., {Anderson}, M.,
  {Neilsen}, D., \& {Motl}, P. 2013, Physical Review Letters, 111, 061105

\bibitem[{{Radice} {et~al.}(2018){Radice}, {Perego}, {Hotokezaka}, {Fromm},
  {Bernuzzi}, \& {Roberts}}]{2018ApJ...869..130R}
{Radice}, D., {Perego}, A., {Hotokezaka}, K., {Fromm}, S.~A., {Bernuzzi}, S.,
  \& {Roberts}, L.~F. 2018, \apj, 869, 130

\bibitem[{{Rubenstein} \& {Schaefer}(2000)}]{2000ApJ...529.1031R}
{Rubenstein}, E.~P., \& {Schaefer}, B.~E. 2000, \apj, 529, 1031

\bibitem[{{Sachdev} {et~al.}(2020){Sachdev}, {Magee}, {Hanna}, {Cannon},
  {Singer}, {SK}, {Mukherjee}, {Caudill}, {Chan}, {Creighton}, {Ewing}, {Fong},
  {Godwin}, {Huxford}, {Kapadia}, {Li}, {Lok Lo}, {Meacher}, {Messick},
  {Mohite}, {Nishizawa}, {Ohta}, {Pace}, {Reza}, {Sathyaprakash}, {Shikauchi},
  {Singh}, {Tsukada}, {Tsuna}, {Tsutsui}, \& {Ueno}}]{2020ApJ...905L..25S}
{Sachdev}, S., {et~al.} 2020, \apjl, 905, L25

\bibitem[{{Schaefer} {et~al.}(2000){Schaefer}, {King}, \&
  {Deliyannis}}]{2000ApJ...529.1026S}
{Schaefer}, B.~E., {King}, J.~R., \& {Deliyannis}, C.~P. 2000, \apj, 529, 1026

\bibitem[{{Shultz} {et~al.}(2015){Shultz}, {Wade}, {Alecian}, \& {BinaMIcS
  Collaboration}}]{2015MNRAS.454L...1S}
{Shultz}, M., {Wade}, G.~A., {Alecian}, E., \& {BinaMIcS Collaboration}. 2015,
  \mnras, 454, L1

\bibitem[{{Tejada Arevalo} {et~al.}(2021){Tejada Arevalo}, {Winn}, \&
  {Anderson}}]{2021arXiv210705759T}
{Tejada Arevalo}, R.~A., {Winn}, J.~N., \& {Anderson}, K.~R. 2021, arXiv
  e-prints, arXiv:2107.05759

\bibitem[{{Turner} {et~al.}(2021){Turner}, {Zarka}, {Grie{\ss}meier}, {Lazio},
  {Cecconi}, {Emilio Enriquez}, {Girard}, {Jayawardhana}, {Lamy}, {Nichols}, \&
  {de Pater}}]{2021A&A...645A..59T}
{Turner}, J.~D., {et~al.} 2021, \aap, 645, A59

\bibitem[{{Vedantham} {et~al.}(2020){Vedantham}, {Callingham}, {Shimwell},
  {Tasse}, {Pope}, {Bedell}, {Snellen}, {Best}, {Hardcastle}, {Haverkorn},
  {Mechev}, {O'Sullivan}, {R{\"o}ttgering}, \& {White}}]{2020NatAs...4..577V}
{Vedantham}, H.~K., {et~al.} 2020, Nature Astronomy, 4, 577

\bibitem[{{Warner}(1983)}]{1983ASSL..101..155W}
{Warner}, B. 1983, in Astrophysics and Space Science Library, Vol. 101, IAU
  Colloq. 72: Cataclysmic Variables and Related Objects, ed. M.~{Livio} \&
  G.~{Shaviv}, 155--171

\bibitem[{{Winn} \& {Fabrycky}(2015)}]{2015ARA&A..53..409W}
{Winn}, J.~N., \& {Fabrycky}, D.~C. 2015, \araa, 53, 409

\end{thebibliography}


\begin{thebibliography}{16}
\expandafter\ifx\csname natexlab\endcsname\relax\def\natexlab#1{#1}\fi

\bibitem[{{Antoine}(2021)}]{2021arXiv210405968A}
{Antoine}, S. 2021, arXiv e-prints, arXiv:2104.05968

\bibitem[{Arnold(1986)}]{Arnold73}
Arnold, V.~I. 1986, Selecta Math. Soviet., 5, 327, selected translations

\bibitem[{{Bildsten} \& {Cutler}(1992)}]{1992ApJ...400..175B}
{Bildsten}, L., \& {Cutler}, C. 1992, \apj, 400, 175

\bibitem[{{Buckley} {et~al.}(2017){Buckley}, {Meintjes}, {Potter}, {Marsh}, \&
  {G{\"a}nsicke}}]{2017NatAs...1E..29B}
{Buckley}, D.~A.~H., {Meintjes}, P.~J., {Potter}, S.~B., {Marsh}, T.~R., \&
  {G{\"a}nsicke}, B.~T. 2017, Nature Astronomy, 1, 0029

\bibitem[{{Goldreich} \& {Julian}(1969)}]{GJ}
{Goldreich}, P., \& {Julian}, W.~H. 1969, \apj, 157, 869

\bibitem[{{Hansen} \& {Lyutikov}(2001)}]{2001MNRAS.322..695H}
{Hansen}, B.~M.~S., \& {Lyutikov}, M. 2001, \mnras, 322, 695

\bibitem[{{Lyutikov}(2011)}]{2011PhRvD..83l4035L}
{Lyutikov}, M. 2011, \prd, 83, 124035

\bibitem[{{Lyutikov}(2019)}]{2019MNRAS.483.2766L}
---. 2019, \mnras, 483, 2766

\bibitem[{Moffatt \& Dormy(2019)}]{moffatt_dormy_2019}
Moffatt, K., \& Dormy, E. 2019, Self-Exciting Fluid Dynamos, Cambridge Texts in
  Applied Mathematics (Cambridge University Press)

\bibitem[{{Most} \& {Philippov}(2020)}]{2020ApJ...893L...6M}
{Most}, E.~R., \& {Philippov}, A.~A. 2020, \apjl, 893, L6

\bibitem[{{Palenzuela} {et~al.}(2013){Palenzuela}, {Lehner}, {Ponce},
  {Liebling}, {Anderson}, {Neilsen}, \& {Motl}}]{2013PhRvL.111f1105P}
{Palenzuela}, C., {Lehner}, L., {Ponce}, M., {Liebling}, S.~L., {Anderson}, M.,
  {Neilsen}, D., \& {Motl}, P. 2013, Physical Review Letters, 111, 061105

\bibitem[{{Petit} {et~al.}(2013){Petit}, {Owocki}, {Wade}, {Cohen},
  {Sundqvist}, {Gagn{\'e}}, {Ma{\'\i}z Apell{\'a}niz}, {Oksala}, {Bohlender},
  {Rivinius}, {Henrichs}, {Alecian}, {Townsend}, {ud-Doula}, \& {MiMeS
  Collaboration}}]{2013MNRAS.429..398P}
{Petit}, V., {et~al.} 2013, \mnras, 429, 398

\bibitem[{{Radice} {et~al.}(2018){Radice}, {Perego}, {Hotokezaka}, {Fromm},
  {Bernuzzi}, \& {Roberts}}]{2018ApJ...869..130R}
{Radice}, D., {Perego}, A., {Hotokezaka}, K., {Fromm}, S.~A., {Bernuzzi}, S.,
  \& {Roberts}, L.~F. 2018, \apj, 869, 130

\bibitem[{{Rubenstein} \& {Schaefer}(2000)}]{2000ApJ...529.1031R}
{Rubenstein}, E.~P., \& {Schaefer}, B.~E. 2000, \apj, 529, 1031

\bibitem[{{Schaefer} {et~al.}(2000){Schaefer}, {King}, \&
  {Deliyannis}}]{2000ApJ...529.1026S}
{Schaefer}, B.~E., {King}, J.~R., \& {Deliyannis}, C.~P. 2000, \apj, 529, 1026

\bibitem[{{Warner}(1983)}]{1983ASSL..101..155W}
{Warner}, B. 1983, in Astrophysics and Space Science Library, Vol. 101, IAU
  Colloq. 72: Cataclysmic Variables and Related Objects, ed. M.~{Livio} \&
  G.~{Shaviv}, 155--171

\end{thebibliography}

\end{document}